\documentclass[aps, superscriptaddress, preprint]{revtex4}
\usepackage{amsmath}
\usepackage{amsfonts}
\usepackage{amssymb}
\usepackage{graphicx}
\usepackage{units}
\usepackage{color}
\usepackage{xspace}
\usepackage[normalem]{ulem}
\usepackage{natbib}
\usepackage{textcomp}
\usepackage{gensymb}
\usepackage{romannum}

\newcommand{\fig}[1]{Fig.~\ref{#1}}

\DeclareUnicodeCharacter{2009}{\,}
\DeclareUnicodeCharacter{2212}{-}

\begin{document}

\title{Dimensional Crossover and Emergence of Novel Phases in Puckered PdSe$_2$
under Pressure}

\author{Tanima Kundu}
\email{tanima.kundu96@gmail.com}
\affiliation{School of Physical Sciences, Indian Association for the Cultivation of Science, 2A $\&$ B
Raja S. C. Mullick Road, Jadavpur, Kolkata - 700032, India
}

\author{Soumik Das}
\affiliation{School of Physical Sciences, Indian Association for the Cultivation of Science, 2A $\&$ B
Raja S. C. Mullick Road, Jadavpur, Kolkata - 700032, India
}

\author{Koushik Dey}
\affiliation{School of Physical Sciences, Indian Association for the Cultivation of Science, 2A $\&$ B
Raja S. C. Mullick Road, Jadavpur, Kolkata - 700032, India
}

\author{Boby Joseph}
\affiliation{Elettra - Sincrotrone Trieste S.C. p. A., S.S. 14, Km 163.5 in Area Science Park, Basovizza 34149, Italy}

\author{Chumki Nayak}
\affiliation{Department of Physical Sciences, Bose Institute, Unified academic campus, EN 80, Sector V, Bidhan Nagar, Kolkata- 700091, India}

\author{Mainak Palit}
\affiliation{School of Physical Sciences, Indian Association for the Cultivation of Science, 2A $\&$ B
Raja S. C. Mullick Road, Jadavpur, Kolkata - 700032, India
}

\author{Sanjoy Kr Mahatha} 
\affiliation{UGC-DAE Consortium for Scientific Research, Khandwa Road, Indore 452001, Madhya Pradesh, India}

\author{Kapildeb Dolui}
\affiliation{Department of Materials Science \& Metallurgy, University of Cambridge, 27 Charles Babbage Road, Cambridge CB3 0FS, United Kingdom}
\affiliation{Department of Physics, Indian Institute of Technology Tirupati, Tirupati, Andhra Pradesh 517619, India}

\author{Subhadeep Datta}
\email{sspsdd@iacs.res.in}
\affiliation{School of Physical Sciences, Indian Association for the Cultivation of Science, 2A $\&$ B
Raja S. C. Mullick Road, Jadavpur, Kolkata - 700032, India
}

\begin{abstract}

\textbf{We investigate the pressure-driven structural and electronic evolution of PdSe\(_2\) using powder X-ray diffraction, Raman spectroscopy, and first-principles calculations. Beyond 2.3 GPa, suppression of the Jahn-Teller distortion induces in-plane lattice expansion and metallization. Around 4.8 GPa, the interlayer \(d_{z^2}-\pi^*\) orbital hybridization drives the dimensional crossover, facilitating the transformation from the 2D distorted to a 3D undistorted pyrite phase. At $\sim$ 9 GPa, a novel phase emerges, characterized by octahedral distortions in the $d$ orbitals of Pd. Structural analysis suggests the presence of marcasite (\(Pnnm\)) or arsenopyrite (\(P2_1/c\)) phase with orthorhombic and monoclinic configurations, respectively. Furthermore, the observed phonon anomaly and electronic structure modifications, including the emergence of flat bands in the high-pressure phases, elucidate the fundamental mechanisms underlying the previously reported exotic superconductivity with an enhanced critical temperature. These results highlight the pivotal role of dimensional crossover and structural transitions in tuning the electronic properties of puckered materials, providing pathways for novel functionalities.}

\end{abstract}

\maketitle

\section{Introduction}
Layered noble metal dichalcogenides (NMDCs) have attracted renewed attention due to their competing capability with unusual electronic properties and diverse exotic phases in the race of van der Waals (vdW) materials \cite{NMDC1}. They exhibit a unique structural motif with the chemical formula $MX_2$ ($M$: noble metals $e.g.$ Pd, Pt and $X$: chalcogenides $e.g.$ S, Se, Te) which can be tuned to a variety of polymorphs under external perturbation (pressure, temperature, etc.) or chemical intercalation \cite{NMDC2,NMDC3,chemmat,PdS2-claudia}. As a viable tuning parameter, pressure plays a crucial role in altering the crystal structure as well as band structure without introducing any chemical disorder. Pressure-induced phase transition and superconductivity especially in vdW materials are more interesting because high pressure reduces the interlayer gap and modify the structural properties accordingly \cite{vdw1,vdw2,vdw3}. In vdW family, Pd$X_2$ ($X$: S, Se) type NMDCs are well-known in recent years for their ultrahigh air-stability and novel anisotropic properties \cite{PdSe2,PdS2}. Depending on the arrangement of dimer anions (X$_2^{2-}$) around the noble metal, they can exhibit several structural polymorphs including orthorhombic, cubic, monoclinic etc. \cite{JAC,soulard,verbeekite1,verbeekite2,rsc}. In ambient condition, both PdS$_2$ and PdSe$_2$ hold orthorhombic 2$O$ phase in a puckered pentagonal morphology. Due to this unique geometry and strong interlayer coupling, such vdW materials possess widely tunable structural and electronic properties \cite{NMDC5}. In particular, pressure-dependent structural study associated with metallization and superconductivity in such materials have gained significant research interests \cite{rsc,PdS2-claudia,claudia,PdS2-2,nanoletter2024}.

In general, transition metal dichalcogenides (TMDCs) exhibit layered or non-layered geometry depending on the coordination between metal $d$ orbitals and chalcogen $p$ orbitals. Non-layered TMDCs like NiSe$_2$, MnS$_2$ \cite{PdS2-claudia,MnS2-1,MnS2-2,MnTe2} etc. usually adopt three-dimensional pyrite (cubic) lattice involving the octahedral configuration of the cationic $d^8$ orbital bonded with six chalcogen dimers. In contrast, layered ambient PdSe$_2$ exhibits square-planar coordination of Pd$^{2+}$ ion with four Se$-$Se dimers along with a vdW interaction between two adjacent layers. The strong spin-orbit coupling in such 4$d$ or 5$d$ noble metal dichalcogenides can lead to exotic electronic structure as well as emerging superconducting states under hydrostatic pressure \cite{claudia,PdS2-claudia}. Despite the similar dimeric arrangement, non-layered NiSe$_2$ with strongly correlated Ni 3$d$ orbitals does not exhibit superconductivity \cite{PdS2-claudia}. On the other hand, layered TMDCs where the vdW force is very weak, there is always a possibility of structural transition $via$ layer sliding \cite{ReSe2-1,MoS2-sliding}, whereas in NMDCs the interlayer coupling is strong enough to resist the layer sliding under compression which may assist the interlayer hybridization resulting in a 2D to 3D dimensional crossover \cite{PtSe2}. However, other NMDCs like PtX$_2$ which do not possess the puckered structure, undergo a pressure-induced dimensional crossover without any trace of structural transition \cite{PtSe2,PtSe2-2}. Hence, the unique puckered geometry coupled with emerging electronic traits in PdSe$_2$ may offer a potential platform in the race of metal dichalcogenides to observe an intriguing interplay between the pressure-modulated structural and electronic phases.

High-pressure powder XRD (HPXRD) as reported by Soulard $et \ al$. \cite{soulard} implies that PdSe$_2$ exhibits orthorhombic structure upto pressure 6 GPa and beyond that it emerges pyrite (cubic $Pa\bar{3}$) phase with Se$-$Se dumbbell of two adjacent layers occupying the vacant states of Pd$^{2+}$ site, hence generates PdSe$_6$ octahedral coordination. The pyrite phase coexists with the ambient phase till 11 GPa. ElGhazali $et \ al$. \cite{claudia} observed the phonon anomaly in PdSe$_2$ $via$ high pressure Raman spectroscopy and comparing the characteristic phonon behavior with the previous XRD study, the phase coexistence is described in between 6$-$10 GPa. Metallization followed by superconductivity are also demonstrated by electrical transport measurements \cite{claudia}. Later, Jiang $et \ al$ \cite{Jiang} claimed that the phase coexistence starts from 6.2 GPa and it continues throughout the experimental pressure range upto 14.8 GPa. Such a wide phase-coexistence window could potentially results from the local pressure-gradient effect which can be reduced by sophisticated instrumentation in order to resolve the pressure-induced phases in a precise manner. Moreover, pyrite TMDCs $e.g.$ MnX$_2$, FeX$_2$ etc. are capable of structural transformation under pressure by altering the effective packing of MX$_6$ octahedra \cite{MnS2-1,MnS2-2,MnTe2,FeS2} which was not experimentally identified in pyrite PdSe$_2$ till now.

Here we focus on a detail structural analysis of puckered PdSe$_2$ using synchrotron powder X-ray diffraction study under hydrostatic pressure. This study is complemented with the high pressure investigation of vibrational Raman modes in order to obtain a comprehensive insight into the structure of PdSe$_2$ in a precise way. Various structural and electronic transitions are identified at different pressure regimes. It was observed that, PdSe$_2$ exhibits in-plane lattice expansion as well as metallization for pressure $>$ 2.3 GPa. In addition, the rapid contraction of the interlayer vdW gap under pressure promotes the structural phase transition into the pyrite phase $via$ dimensional crossover. Notably, the local pressure gradient effect for which the phase-coexistence could persist in a long range as reported previously, gets significantly reduced in our measurements. A further transformation to a new phase was manifested at higher pressure ($\sim$ 9 GPa) which was not explored previously. Two possible phases, marcasite and arsenopyrite, comprising orthorhombic ($Pnnm$) and monoclinic ($P2_1/c$) structural motif, respectively were identified by XRD refinement, which arise from the pressure-induced octahedral distortion in the antecedent pyrite phase. Anomalous behavior of the phonon modes indicates the electron redistribution and evident structural phase transition. Furthermore, the thermodynamic stability of the high- pressure phases was studied by the enthalpy calculations using first-principle density functional theory. Orbital-projected band structures were depicted to highlight the contribution of Pd $d$- orbitals, providing insights into the octahedral splitting and metallicity across various high-pressure phases. Our work thus encompasses a complete study incorporating emergent structural phases in puckered pentagonal PdSe$_2$ accompanied by dimensional crossover, which is quite improbable within the family of two-dimensional transition metal dichalcogenides.

\section{Experimental details}
\label{subsec:exp}
{\bfseries Sample synthesis and phase characterization:} Standard self-flux method was adopted to grow bulk PdSe$_2$ single crystals as described in ref. \cite{chemmat}. Ambient powder X-ray diffraction (PXRD) data was collected using a Rigaku SmartLab (Cu K$_\alpha$ radiation, $\lambda$ = 1.5406 \AA) diffractometer to check the phase quality of the well ground as-grown crystals. Single crystal XRD and energy dispersive study were also carried out for structural solution and stoichiometric characterization, respectively as reported in ref. \cite{chemmat}.

{\bfseries High-pressure synchrotron X-ray diffraction:} A bunch of PdSe$_2$ crystals were well ground and pressed into a pellet. High pressure X-ray diffraction measurements were carried out at the XPRESS beamline of the ELETTRA synchrotron, Trieste, Italy \cite{elettra}. Sample along with a Ruby ball, and pressure transmitting medium (4:1 mixture of methanol and ethanol) were loaded in the 100-µm hole of a gasketed membrane-driven diamond anvil cell (DAC) having a diamond culet size of 400 µm. Pressure value within the DAC was monitored by the fluorescence spectra of Ruby. Monochromatic X-ray beam of wavelength 0.4959 \AA, and diameter 50 µm was utilized for the high-resolution X-ray diffraction measurements. The sample-to-detector distance was calibrated using the X-ray diffraction pattern of LaB$_6$. A large area PILATUS3 S 6M detector, placed at a distance of 940 mm from the sample position, was used to collect the diffraction data. Two-dimensional (2D) diffraction rings obtained from the image plate detector were converted to intensity versus two theta plots using DIOPTAS software \cite{dioptus}. Finally, complete structural parameters and the evolution of the bond lengths with pressure were extracted using the FULLPROF software package \cite{fullprof}. 

{\bfseries High-pressure Raman spectroscopy:} The high-pressure Raman measurement was carried out in a diamond anvil cell (DAC) from DAC Tools with a culet diameter of 350 $\mu$m. Bulk PdSe$_2$ flakes were mechanically transferred using PDMS from the crystal onto the diamond anvil using a micromanipulator under an optical microscope \cite{feps3}. Instead of directly loading the bulk crystal, a targeted transfer technique is used to prevent the unintended mixing of Se flux with the crystal. A stainless steel gasket was used to form the sample chamber, which was pre-indented to a thickness of 60 $\mu$m, with a 150-$\mu$m diameter hole drilled at the center of the indentation. Ruby spheres were placed in the chamber alongside the sample to measure the pressure during the experiment. The chamber was filled with a 4:1 methanol-ethanol mixture, which acted as the hydrostatic pressure-transmitting medium (PTM). Raman spectra were acquired using a micro-Raman setup equipped with a spectrometer (LabRAM HR, Jobin Yvon) and a Peltier-cooled CCD detector. An air-cooled argon-ion laser with a wavelength of 488 nm served as the excitation source, and the beam was focused on the sample using a 50x long-working-distance objective lens with a numerical aperture of 0.50. A constant laser power of about 200 $\mu$W was maintained to prevent sample damage throughout the experiment.

\section{Computational details}
\label{subsec:theory}

To explore possible crystal structures at various pressures, we performed energetic calculations using density functional theory for different crystalline bulk phases of PdSe$_2$ listed in the \textsc{OQMD}~\cite{db_oqmd}, and Materials Project database~\cite{db_mp}. Geometry optimizations and all subsequent energetic calculations performed using \textsc{castep} \cite{castep} used the RSCAN functional~\cite{rscan}, a 600 eV plane-wave cutoff, a $k$-point spacing of 2$\pi\times$ 0.03 \AA$^{-1}$, and default \textsc{castep} C19 pseudopotentials. The Tkatchenko-Scheffler (TS) based many-body dispersion scheme (MBD) is used for long-range dispersion correction~\cite{ts-mbd1, ts-mbd2}. The structures were converged to give forces and stresses within 0.01 eV/\AA \ and 0.01 GPa.

For orbital-projected band structure and electron localization function (ELF) calculations density functional theory calculations were carried out using the Vienna Ab initio Simulation Package (VASP) \cite{VASP1,VASP2} with the Perdew-Burke-Ernzerhof (PBE) exchange-correlation functional \cite{PBE}. In order to process the orbital-projected band structures, VASPKIT code was used \cite{vaspkit}. To visualize the crystal structures and ELF, VESTA software \cite{VESTA} was used.

Electron-phonon coupling calculations were carried out within the framework of density functional perturbation theory (DFPT) using Quantum Espresso \cite{dfpt,qe}. The PBE exchange-correlation functional was employed in combination with scalar-relativistic ultrasoft Vanderbilt pseudopotentials. Energy cutoffs for the plane-wave basis and charge density were set to 60 Ry and 480 Ry, respectively. A q-mesh of 3×3×3, 3×3×3, and 3×3×4 and a k-mesh of 9 × 9 × 9, 9 × 9 × 9, and 12 × 12 × 16 were used for the $P2_1/c$, $Pnnm$, and $Pa\bar{3}$ phases of PdSe$_2$ at 12 GPa, respectively. A phonon self-convergence threshold of 10$^{−16}$ was adopted for the calculation of electron-phonon coupling elements. The superconducting transition temperature $T_c$ was estimated using the Allen–Dynes modified McMillan equation \cite{McMillan}, with an effective screened Coulomb repulsion constant $\mu^\star$ = 0.1. Dirac $\delta$-functions for electrons and phonons were approximated by smearing functions with widths of 0.03 Ry and 0.3 mRy, respectively.

\section{Results and discussion}
\label{sec:results_discussions}

Bulk PdSe$_2$ is a semiconducting NMDC having unique pentagonal structure in 2D layers. It crystallizes in the orthorhombic $Pbca$ space group as shown in Fig. S1(a) of the Supplemental Material (SM) \cite{supple}. Fig. S1(b) of SM \cite{supple} shows the Rietveld refined synchrotron powder X-ray diffraction data measured at ambient condition, which reveals the orthorhombic $Pbca$ phase of PdSe$_2$ mixed with $\sim$ 1.4\% Se due to the self-flux growth mechanism \cite{pxrd-adfm}. The lattice parameters obtained from ambient PXRD data are $a=5.74$ {\AA}, $b=5.86$ {\AA}, $c=7.69$ {\AA} and $\alpha=\beta=\gamma=90\degree$ with an interlayer van der Waals spacing along the $c$-axis. Unit cell of PdSe$_2$ consists of four Pd atoms, each one is tetra-coordinated with the chalcogen dimers in square planar geometry due to the typical $d^8$ configuration of Pd$^{2+}$ ions and forms a unique puckered pentagonal morphology.

\begin{figure}[ht!]
\centerline{\includegraphics[scale=0.47,clip]{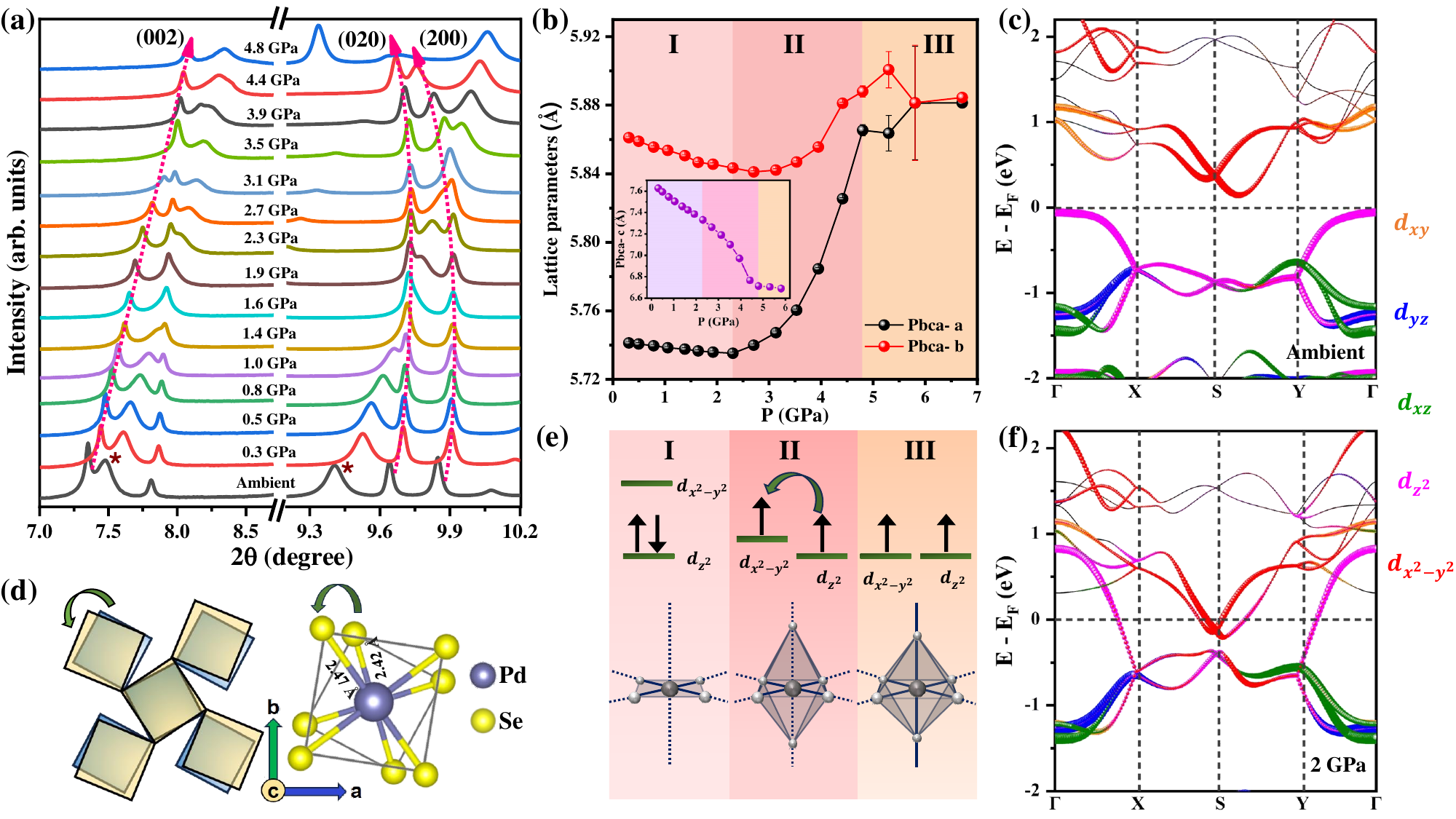}}
\caption{Structural and electronic evolution at low pressure regime: (a) Evolution of the Bragg peaks corresponding to the lattice planes (200), (020), and (002). (200) and (020) peaks associated with the in-plane lattice are observed to exhibit an anomalous shift towards lower 2$\theta$ under compression above 2.3 GPa. The peaks marked by * at ambient correspond to the diffraction lines of Se- flux. (b) Pressure evolution of the in-plane lattice parameters $a$ and $b$ segmented in three regimes according to the structural rearrangement and electronic transition. The regime I depicts the usual compression of the lattice axes corresponding to the ambient distorted pyrite structure. Regime II reveals the in-plane lattice expansion and regime III portrays the persistence of the ambient phase due to local presssure gradient. Inset shows the rapid reduction of the vdW lattice axis $c$ under pressure with a certain change in slope at 2.3 GPa. (c) Orbital-projected semiconducting band structure of ambient PdSe$_2$ depicting the valence band maxima contributed by Pd $d_{z^2}$ orbitals and conduction band minima contributed by $d_{x^2-y^2}$  orbitals. (d) Schematic illustration of PdSe$_4$ square-planar expansion and rotation (left), accompanied by the elongation of Pd-Se bond lengths (right) beyond 2.3 GPa. (e) Regime I: Jahn-Teller distorted spin-singlet square-planar configuration of ambient PdSe$_2$ with fully occupied $d_{z^2}$ orbital. Regime II: Commencement of intralayer charge transfer from Pd $d_{z^2}$ to $d_{x^2-y^2}$ orbitals above 2.3 GPa reveals the suppression of J-T distortion. Regime III: Equal charge distribution between the degenerate $e_g$ orbitals of Pd comprising of a spin-triplet state and pure octahedral coordination. (f) Population of $d_{x^2-y^2}$ orbitals gradually exhibits metallicity manifested by the metallic band structure at 2 GPa.
\label{Pbca}}
\end{figure}

In order to observe the pressure-induced structural modulation and thermodynamic stability, we performed high pressure PXRD experiments using synchrotron radiation source upto 13.1 GPa at room temperature as illustrated in Fig. S2 of SM \cite{supple}. Under compression upto 2.3 GPa, a gradual shift of all the Bragg peaks was observed towards higher angles in usual manner. However, beyond this pressure regime, the Bragg peaks associated with the lattice planes perpendicular to the vdW direction started shifting towards lower $2\theta$. As shown in \fig{Pbca}(a), the zoomed-in view of some selected reflections clearly depicts that the (002) peak has a typical shifting towards higher $2\theta$ with increasing pressure contrary to the (200) and (020) Bragg peaks which exhibit an anomalous shift towards lower $2\theta$ above 2.3 GPa (see Fig. S3 of SM \cite{supple} for $2\theta$ $vs.$ pressure plots for (200) and (020) peaks). In accordance with the different structural and electronic transitions, pressure-dependence of the lattice parameters of the $Pbca$ phase is segmented in three regimes (I, II, and III) as shown in \fig{Pbca}(b). In regime I, all the lattice axes render a usual decrease under compression. Then an anomalous expansion of the in-plane lattice parameters ``$a$'' and ``$b$'' and a change in slope in the vdW axis ``$c$'' are observed with pressure beyond 2.3 GPa as depicted in regime II. 
In principle, the ambient phase of PdSe$_2$ can be described as a distorted pyrite structure driven by the Jahn-Teller (JT) effect \cite{JT} with the elongation of PdSe$_6$ octahedra along the $c$- axis manifesting a vdW geometry. In an undistorted octahedral environment, the lowest energy state of Pd$^{2+}$ ($d^8$) cation, as dictated by Hund's rule, is a spin-triplet. This state owing to the orbital degeneracy, is not susceptible to the Jahn-Teller (JT) effect. However, if the energy difference between the undistorted ground and distorted excited states with different spin multiplicity is small enough, and the coupling to JT vibrations is sufficiently strong, the system can overcome the energy barrier, leading to the distorted octahedral coordination with a singlet electronic configuration becoming more stable \cite{PdS2-2,soulard}. Consequently, the lower $d_{z^2}$ orbitals get electron-filled and conversely $d_{x^2-y^2}$ orbitals become electron-vacant, thus exhibits a semiconducting characteristics as well as diamagnetic nature of ambient PdSe$_2$. As observed from the orbital-projected band structure at ambient (\fig{Pbca}(c)), the valence band maxima and conduction band minima are predominantly contributed by Pd $d_{z^2}$ and $d_{x^2-y^2}$ orbitals, respectively. Henceforth, the semiconducting distorted pyrite $Pbca$ phase remain energetically favorable at ambient and it persists upto 2.3 GPa justifying the usual decrease of in-plane lattice axes in regime I.

\begin{figure}[ht!]
\centerline{\includegraphics[scale=0.52, clip]{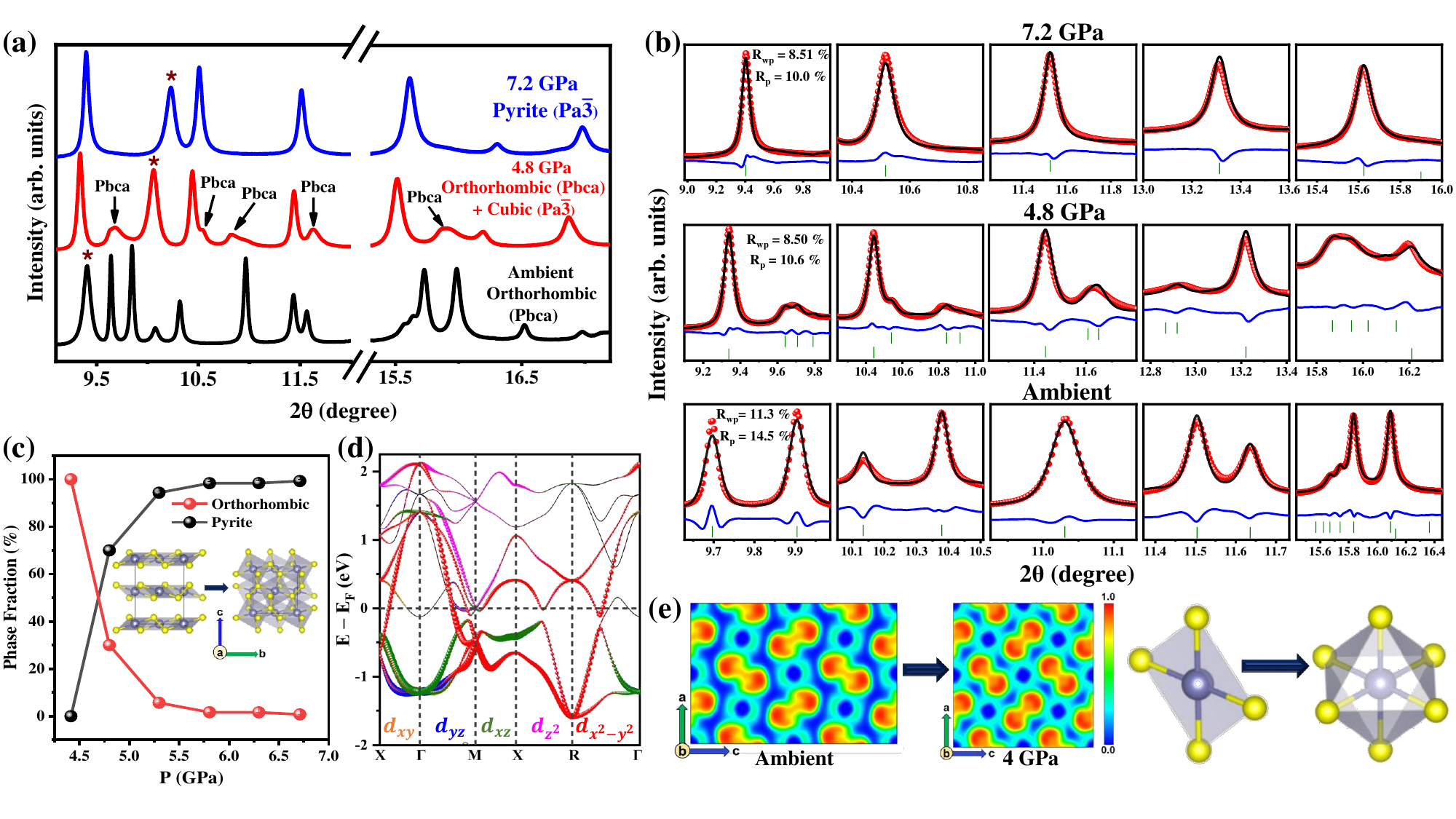}}
\caption{Distorted to undistorted phase transition: (a) Stack plot of the Bragg peaks corresponding to the synchrotron powder XRD at ambient, 4.8 and 7.2 GPa. The * peaks are associated with Se- flux. The ambient phase persists as small humps at 4.8 GPa along with a rapid evolution of the new undistorted pyrite phase. The phase coexistence retains below 7.2 GPa followed by the complete structural transition to pure pyrite phase. (b) Zoomed-in view of some Rietveld refined diffraction lines with reasonably good fitting parameters at the above-mentioned pressure regions. At ambient, the peaks are well-indexed with the orthorhombic $Pbca$ phase. At 4.8 GPa, both $Pbca$ (top Bragg positions) and $Pa\bar{3}$ (bottom Bragg positions) phases are required to fit the XRD pattern revealing the phase coexistence and at 7.2 GPa, all the peaks can be refined well with the pyrite ($Pa\bar{3}$) phase only. (c) Relative phase fraction $vs.$ pressure plot represents the significant reduction of the ambient phase at 4.8 GPa with $<$ 30$\%$ phase fraction and a rapid conversion into the pyrite phase with a phase coexistence window of $\sim$ 2 GPa. (d) Orbital-projected metallic band structure of pyrite PdSe$_2$ where the bands near Fermi surface are predominantly contributed by mixed-filled Pd $d_{x^2-y^2}$ and $d_z^2$ orbitals. (e) Electron localization function (ELF) manifests the interlayer hybridization of Pd $d_z^2$ orbitals with the Se $p$ orbitals from adjacent layers, thereby leading a dimensional crossover and promotes the orthorhombic to pyrite phase transition. In pyrite phase, the square planar PdSe$_4$ unit transforms into an octahedral PdSe$_6$ unit. 
\label{pyrite}}
\end{figure}

With further compression, the intralayer Pd$-$Se bond lengths start increasing (Fig. S4(a) of SM \cite{supple}) leading to an in-plane lattice expansion schematically shown by the elongation as well as rotation of the PdSe$_4$ square-planar units in \fig{Pbca}(d) (Pd$-$Se bond length is 2.42 Å at 2.3 GPa and 2.47 Å at 4.4 GPa). The drastic decrease in interlayer distance leads to a strong overlap repulsion between the already-filled $d_{z^2}$ orbitals of Pd and the antibonding orbitals of Se from the adjacent layers. This initializes charge transfer from $d_{z^2}$ to $d_{x^2-y^2}$ orbital reducing the corresponding energy gap as well as structural distortion and leads to a gradual transformation from singlet to triplet electronic state modifying the Lennard-Jones potential accordingly (shown schematically in \fig{Pbca}(e)). The commencement of semiconducting to metallic transition (from regime I to II) is driven by the strong hybridization of the populated $d_{x^2-y^2}$ orbitals of Pd with chalcogen $p$ orbitals, forming the conduction bands as represented by the projected band structure of distorted PdSe$_2$ at 2 GPa (\fig{Pbca}(f)). At the onset of intralayer singlet to triplet transition, the interlayer hybridization between vacant $d_{z^2}$ of Pd and Se$-\pi^*$ orbitals of the adjacent layers is commenced owing to the significant reduction along the layer direction. The interlayer electronic distribution drastically increases upto 4.8 GPa and gives rise to the undistorted pyrite phase ($Pa\bar{3}$) through a 2D to 3D network type dimensional crossover. Such phase transition can be assigned as a displacive type phase transition in which the low-pressure phase is a distorted form of the high pressure phase and the transition is accomplished by small displacements of the atoms altering the corresponding coordination and bonding \cite{dove-book}. However, the distorted ambient phase coexists with the undistorted pyrite phase upto $\sim$ 7 GPa for which the increment in the lattice parameters associated with the ambient phase persists upto regime III in \fig{Pbca}(b). As shown in \fig{pyrite}(a), the stacked XRD plots clearly indicates that new diffraction peaks appear corresponding to the pyrite phase at 4.8 GPa and the ambient phase persists as small humps which get vanished completely at 7.2 GPa. \fig{pyrite}(b) depicts the zoomed-in view of some refined Bragg peaks at different pressures with reasonably good fitting parameters along with a clear signature of orthorhombic to pyrite phase transition (see supplemental material \cite{supple} for full XRD refined spectra and detail refinement parameters). \fig{pyrite}(c) shows the relative phase fraction $vs.$ pressure plot where it manifests that the distorted pyrite structure is preserved below 4.8 GPa, then both the distorted and undistorted phases coexist upto $\sim$ 7 GPa followed by the evolution of pure pyrite phase from 7.2 GPa. In the pyrite phase, $d_{x^2-y^2}$ orbital dominates near Fermi surface as interpreted by the projected band structure shown in \fig{pyrite}(d) which certainly exhibits a metallic behavior. Furthermore, the electron localization functions (ELF) are mapped in \fig{pyrite}(e) at ambient and 4 GPa to account the interlayer electron distribution which leads to the conversion of square planar PdSe$_4$ unit into a six-coordinated octahedral geometry justifying the dimensional crossover as well. Notably, the pyrite phase unveils paramagnetic nature in contrast to the ambient diamagnetic phase \cite{dia}.

\begin{figure}[ht!]
\centerline{\includegraphics[scale=0.55, clip]{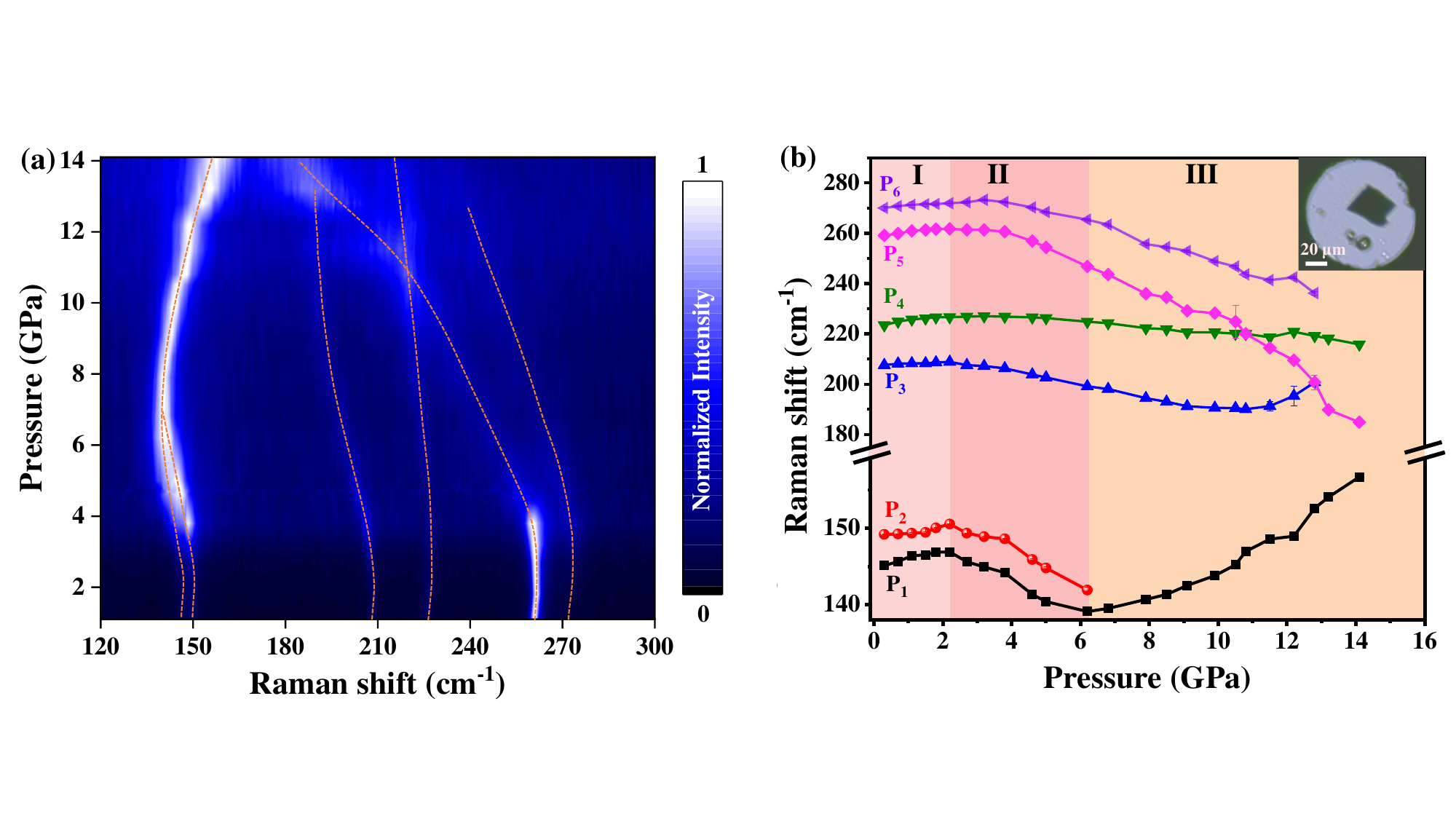}}
\caption{Pressure evolution of phonon modes: (a) False-color map of the Raman scattering response showing the evolution of different phonon modes with pressure. The dashed lines serve as guides to the eye showcasing the progression of each mode. (b) Peak positions of the characteristics phonon modes are plotted against pressure. The invisible error bars are smaller than the data points. In zone I, all phonon modes exhibit a typical blue shift under compression. Zone II shows anomalous softening of the P$_1$ and P$_2$ modes, attributed to intralayer Pd–Se bond stretching and electron transfer involving Pd $d$ orbitals, along with in-plane lattice expansion. This softening persists up to $\sim$ 6.2 GPa, marking the end of Zone II. In Zone III, the merged P$_1$ and P$_2$ modes harden and correspond to the doubly degenerated $E_g$ mode of the undistorted pyrite phase. Meanwhile, the P$_5$ mode, related to Se$–$Se dimer vibrations, begins to soften gradually in zone II and continues to soften rapidly up to the experimentally accessible pressure regime. Inset shows the representative optical image of bulk PdSe$_2$ flake loaded in the anvil cell along with Ruby balls.       
\label{Raman}}
\end{figure}

In order to examine the pressure evolution of characteristics phonon modes, high pressure Raman spectroscopy was performed. At ambient, PdSe$_2$ exhibits six Raman modes (3$A_g$ and 3$B_{1g}$) \cite{PdSe2,chemmat}. To systematically investigate the pressure-induced phonon behavior in conjunction with structural transformations observed in present XRD measurements supported by plausible electronic transitions, the color-map of Raman scattering response is illustrated against pressure as shown in \fig{Raman}(a) and the peak positions of different Raman modes under pressure are plotted in \fig{Raman}(b) (see Fig. S5(a) of SM \cite{supple} for stacked Raman spectra at different pressures). Initially, all the phonon modes show typical blue shift under compression in zone I analogous to XRD-identified regime I and exhibit an anomalous softening thereafter. In zone II, P$_1$ and its adjacent P$_2$ modes (librational $A_g^1$ and $B_{1g}^1$ modes) exhibit a clear red shift resulting from the stretching vibration of intralayer Pd$-$Se bonds beyond 2.3 GPa (Fig. S4(a) of SM \cite{supple}), which is associated with the electron transfer between Pd $d$ orbitals and in-plane lattice expansion (\fig{Pbca}(b)). Another significant mode P$_5$ ($A_g^3$) begins to soften gradually at the onset of zone II. The softening becomes more pronounced afterwards caused by the increase in Se$-$Se dimer bond length (Fig. S4(b) of SM \cite{supple}). This effect is attributed by the Se$_2^{2-}$ dimer consuming more electrons by occupying the anti-bonding states due to interlayer hybridization with a rapid reduction of interlayer spacing under compression. P$_5$ remains the most prominent mode up to $\sim$ 4 GPa as seen from the color map in \fig{Raman}(a), characteristics of the distorted pyrite phase. Beyond this pressure, P$_1$ and P$_2$ attain prominence implying the onset of undistorted pyrite phase, which emerges around 4.8 GPa according to XRD. However, P$_1$ and P$_2$ remain discernible separately up to 6.2 GPa (marked as the termination point of zone II) consistent with the phase coexistence regime as evident from XRD (\fig{pyrite}(c)). With further compression (zone III), P$_1$ and P$_2$ merge into a doubly degenerated $E_g$ mode, reflecting the symmetry of the pure pyrite phase. This $E_g$ mode hardens in usual manner upto the experimentally accessible pressure range with a regular contraction of Pd$-$Se bonds under pressure.     
The full width at half maxima (FWHM) of the phonon modes are graphed against pressure, as shown in Fig. S6 of SM \cite{supple}. The significant broadening of the FWHM in the P$_5$ mode from 2.7 GPa to 5 GPa is associated with the transition from distorted pyrite to pyrite phase where it becomes subservient, leading to regular behavior thereafter. Notably, substantial FWHM broadening in the P$_1$ mode is observed beyond $\sim$ 9 GPa. Certain instances of FWHM broadening may intimate the bifurcation of this mode with the emergence of a low-symmetric phase. Other two modes, P$_3$ and P$_4$, also exhibit similar FWHM broadening at this regime, signifying a plausible structural rearrangement. Although anhydrostatic pressure condition at higher pressure may be a possible issue in FWHM broadening, still in case of superconducting phases phonon broadening is a typical phenomena due to high electron-phonon coupling, which further indicates the possible emergence of new structural phases.

\begin{figure}[ht!]
\centerline{\includegraphics[scale=0.6, clip]{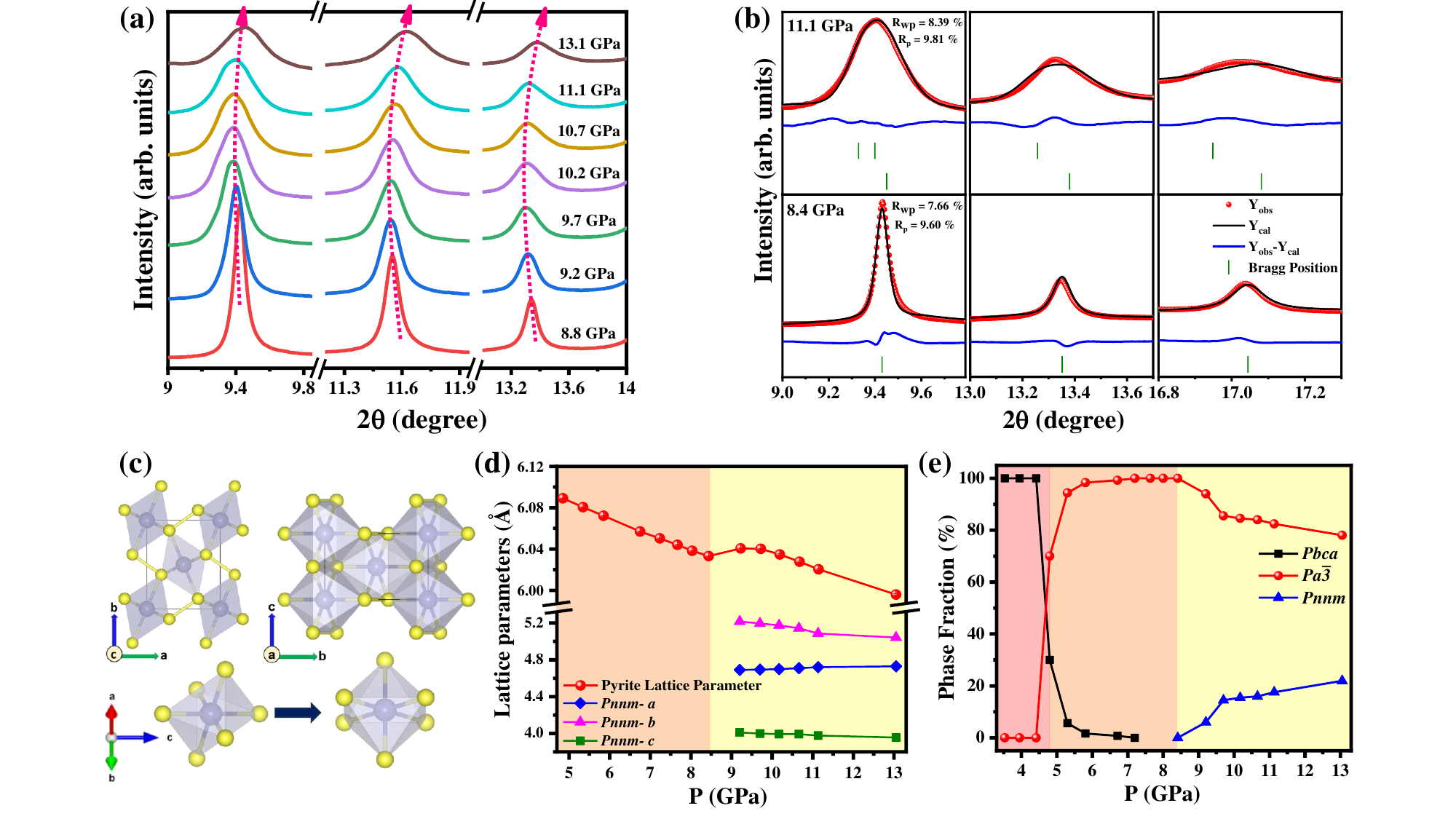}}
\caption{Structural evolution at high pressure: (a) Anomalous evolution of certain Bragg peaks towards lower angles indicating a structural phase transition. (b) Zoomed-in view of some refined Bragg peaks are represented which are well-fitted with the pyrite phase at 8.4 GPa. At 11.1 GPa, a new orthorhombic phase (top Bragg positions), marcasite (space group $Pnnm$) is introduced to refine the diffraction peaks along with the pyrite phase (bottom Bragg positions). (c) Different orientations of the marcasite structure in the upper panel and the octahedral rotation from pyrite to marcasite transition in the lower panel are interpreted schematically. (d) Lattice parameters for both pyrite and marcasite phases are plotted against pressure. (e) Relative phase fraction $vs.$ pressure plot including all the pressure regimes. A maximum 22\% phase fraction for the new marcasite phase is obtained upto the experimentally accessible pressure regime.  
\label{marcasite}}
\end{figure}

Interestingly, beyond $\sim$ 9 GPa, an anomalous shift of certain Bragg peaks towards lower $2\theta$ is also detected in the XRD pattern (\fig{marcasite}(a)). In this regime the Rietveld refinement suggests the existence of two possible phases, marcasite ($Pnnm$) and arsenopyrite ($P2_1/c$), which can coexist with the primary pyrite phase. While symmetry-lowering structural transition often activate new Raman modes owing to relaxed selection rule, no new phonon mode is detected within the experimentally accessible pressure regime and spectral resolution of the spectrometer in this study. The structures of pyrite, marcasite, and arsenopyrite bear close kinship, as reported in prior studies \cite{jpcs,jpcc}. In each, a cation is enveloped by six anions at the vertices of a slightly distorted octahedron, while each anion resides in a distorted tetrahedral environment, forming anion dimers. In pyrite, adjacent octahedra share a single vertex. Conversely, in marcasite and arsenopyrite, neighboring octahedra share an edge, resulting in linear cation chains. These chains exhibit uniform spacing in marcasite but alternate between short and long links in arsenopyrite. Transition between these phases in PdSe$_2$ may stem from reorientation of anionic dimers and uneven compression along lattice axes, possibly restricting the appearance of any new mode \cite{jpcc}. Moreover, new phases, coexisting with the pyrite phase in minor fractions up to the accessible pressure limit, may potentially obscure the detection of new modes. The high-pressure phases, being purely metallic, further complicate the spectral identification of subtle phononic signatures.

Considering the marcasite phase, the zoomed-in view of some refined Bragg peaks are plotted against pressure in \fig{marcasite}(b) which signifies the coexistence of the new phase along with the pyrite phase, thereby causes asymmetric broadening and doublet nature of the corresponding Bragg peaks. Consequently, the corner sharing octahedra in the pyrite phase start rotating under compression and generates an edge-sharing octahedral configuration with a larger $b$ parameter than $a$ and $c$ as shown in \fig{marcasite}(c), thus recognized as B- marcasite structure \cite{FeSe2,MnS2-2} ($a$= 4.694 \AA, $b$= 5.194 \AA, $c$= 3.999 \AA \ at 9.7 GPa). Both pyrite and marcasite lattice parameters are plotted at different pressure in \fig{marcasite}(d). The slight lattice expansion in the cubic phase may be due to the structural reconstruction in the transition regime. A maximum 22\% phase fraction of the marcasite phase is obtained till the experimentally accessible pressure (\fig{marcasite}(e)) (refinement details are provided in the supplementary section \cite{supple}).

\begin{figure}[ht!]
\centerline{\includegraphics[scale=0.6, clip]{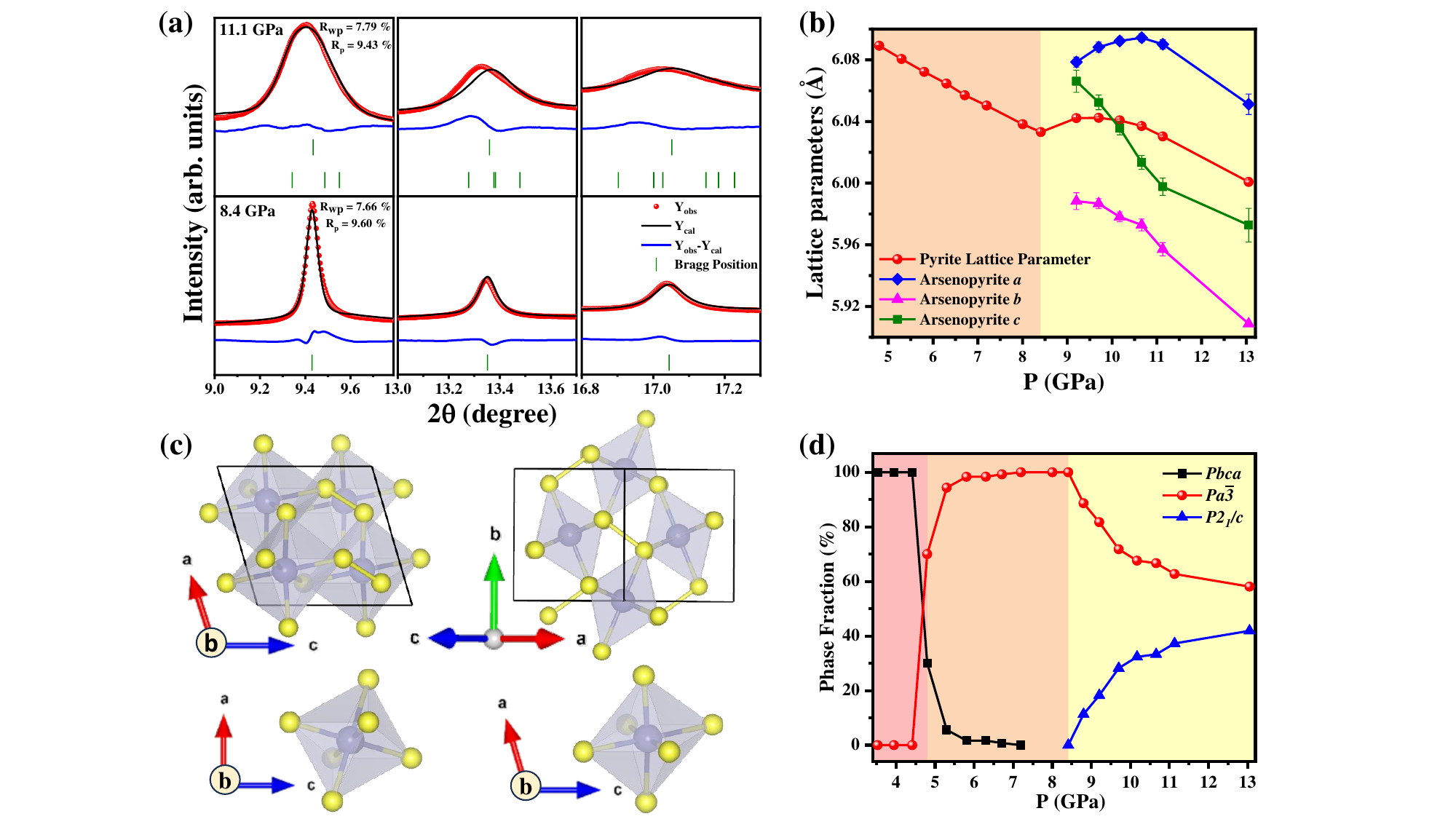}}
\caption{Structural evolution at high pressure: (a) Zoomed-in view of some refined Bragg peaks are represented which are well-fitted with the pyrite phase at 8.4 GPa. At 11.1 GPa, those peaks are refined with another possible new phase (bottom Bragg positions), monoclinic arsenopyrite (space group $P2_1/c$) along with the pyrite phase (top Bragg positions). (b) Evolution of lattice parameters for both pyrite and arsenopyrite phases are plotted against pressure. (c) Different orientations of the arsenopyrite structure in the upper panel and the octahedral rotation from pyrite to arsenopyrite transition in the lower panel are interpreted schematically. (d) Relative phase fraction $vs.$ pressure plot including all the pressure regimes. A maximum 42\% phase fraction is obtained for the arsenopyrite phase upto the experimentally accessible pressure regime. 
\label{arsenopyrite}}
\end{figure}

Another possible phase with monoclinic space group $P2_1/c$ is designated by the name arsenopyrite. The new diffraction lines can be well indexed to the arsenopyrite structure yielding a well-refined fitting parameters as shown in \fig{arsenopyrite}(a). The lattice parameters corresponding to pyrite and arsenopyrite phases are plotted against pressure as shown in \fig{arsenopyrite}(b). Variation of the lattice angle $\beta$ under pressure for the monoclinic phase is depicted in Fig. S12(a) of SM \cite{supple}. The name arsenopyrite basically represents the structure of FeAsS in which `As' doping in FeS$_2$ creates a chemical pressure within the inherent structure and thereby generates a monoclinic architecture \cite{FeAsS-1,FeAsS-2}. In pyrite PdSe$_2$, hydrostatic pressure can generate such type of monoclinic geometry correlated with the orientation and bond-strength of Se$–$Se dumbbells, and the distortion of PdSe$_6$ octahedrons on the $ac$ plane, which results in the larger values of lattice parameters $a$ and $c$ than that of $b$ \cite{MnSe2} ($a$= 6.088 \AA, $b$= 5.986 \AA, $c$= 6.052 \AA \ and $\beta$= 90.6$\degree$ at 9.7 GPa) as portrayed by the structural view in \fig{arsenopyrite}(c). A maximum 42$\%$ phase fraction of the arsenopyrite phase is obtained within the experimentally accessible pressure region (\fig{arsenopyrite}(d)) (refinement details are provided in the supplementary section \cite{supple}).

The variations of unit-cell volume per formula unit for the four phases under sequential pressure-induced structural transition are depicted in Fig. S12(b) of SM \cite{supple}. At 4.8 GPa, the relative volume collapse from ambient orthorhombic to pyrite phase ($\Delta{V}/V_{Pbca}$) is about 2.6\%, whereas at 9.2 GPa, the relative volume collapse from pyrite to marcasite or arsenopyrite phase ($\Delta{V}/V_{Pa\bar{3}}$) are 11.1\% and 0.2\%, respectively. 

To validate the new structural phase transition, enthalpy calculation was carried out using density functional theory (DFT) incorporating all the bulk phases of PdSe$_2$ from ambient to 20 GPa. At ambient pressure, the orthorhombic $Pbca$ phase is found to be the most stable. We subsequently shortlisted structures with energy differences below 300 meV per formula unit (f.u.) relative to the ambient $Pbca$ phase within the pressure range of 0 to 20 GPa. Simulated XRD patterns for these selective structures in the experimentally accessible pressure ranges identified that the orthorhombic $Pbca$, cubic $Pa\bar{3}$, orthorhombic $Pnnm$, and monoclinic $P2_1/c$ phases closely match with the experimental XRD as shown by the Rietveld refined diffraction patterns at different pressure regimes in Fig. S7 to S11 of SM \cite{supple}. Additionally, our zone-center based phonon calculations confirms that all these four phases are dynamically stable across the investigated pressure range. 

\begin{figure}[ht!]
\centerline{\includegraphics[scale=0.55, clip]{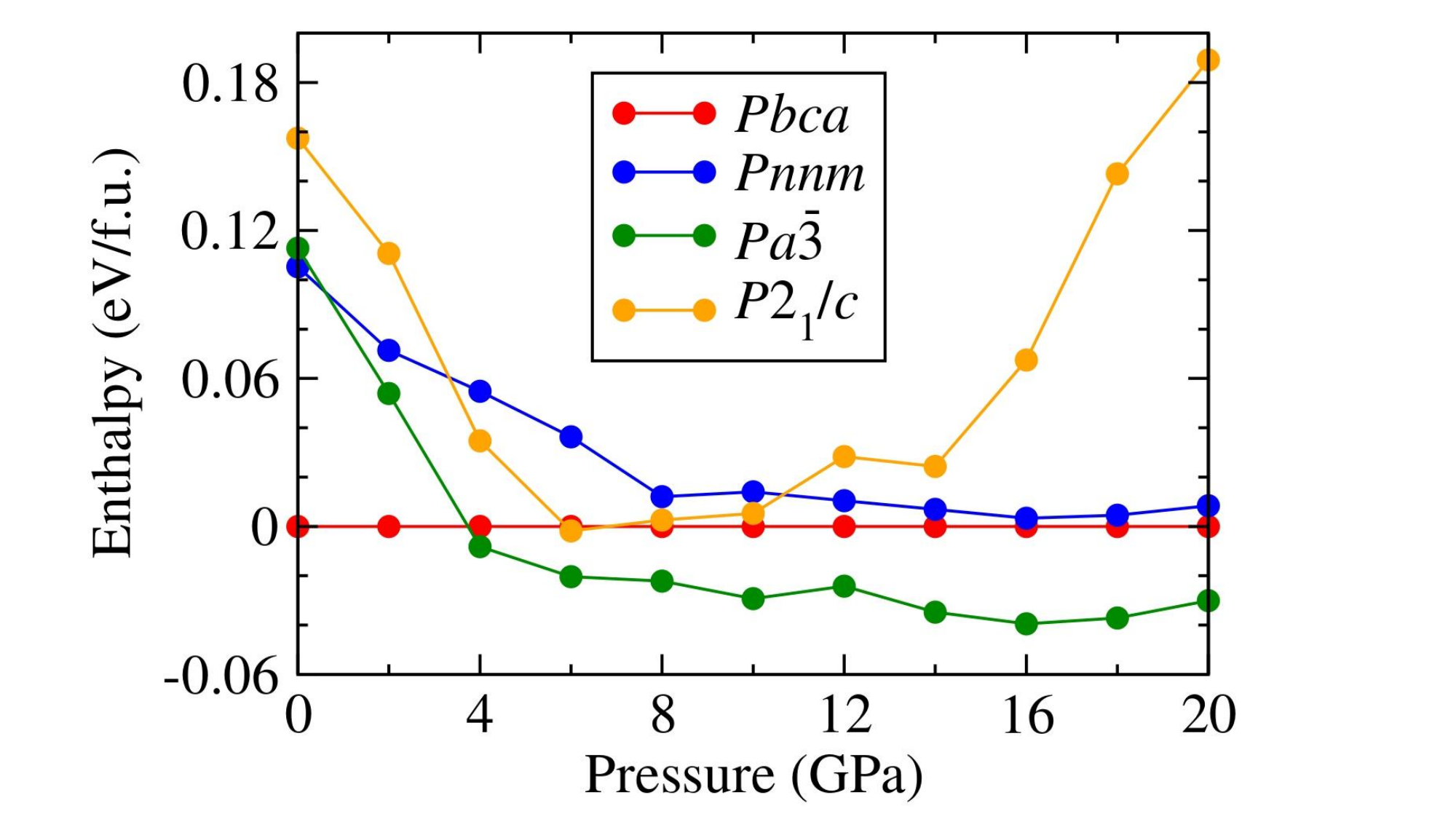}}
\caption{Calculated enthalpy formation energies per formula unit (f.u.) at various pressures for different phases, $Pa\bar{3}$ (green), $Pnnm$ (blue), and $P2_1/c$ (orange) relative to the ambient $Pbca$ phase (red line) of PdSe$_2$ obtained using DFT with the RSCAN functional incorporating many-body dispersion corrections.
\label{enthalpy}}
\end{figure}

The enthalpy formation energies of the four phases, shown in \fig{enthalpy}, indicate that the ambient-pressure $Pbca$ phase transforms into the pyrite ($Pa\bar{3}$) phase around 4 GPa, which remains the lowest-energy state up to 20 GPa. Interestingly, two other phases, namely $P2_1/c$ and $Pnnm$ are energetically close ($<$ 80 meV/f.u.) to the $Pa\bar{3}$ phase in the pressure range 6$-$14 GPa and 8$-$20 GPa, respectively. As the enthalpy calculations do not include temperature effects, thermal fluctuations at room temperature, it could potentially enable access to these energetically close phases.
  
\begin{figure}[ht!]
\centerline{\includegraphics[scale=0.55, clip]{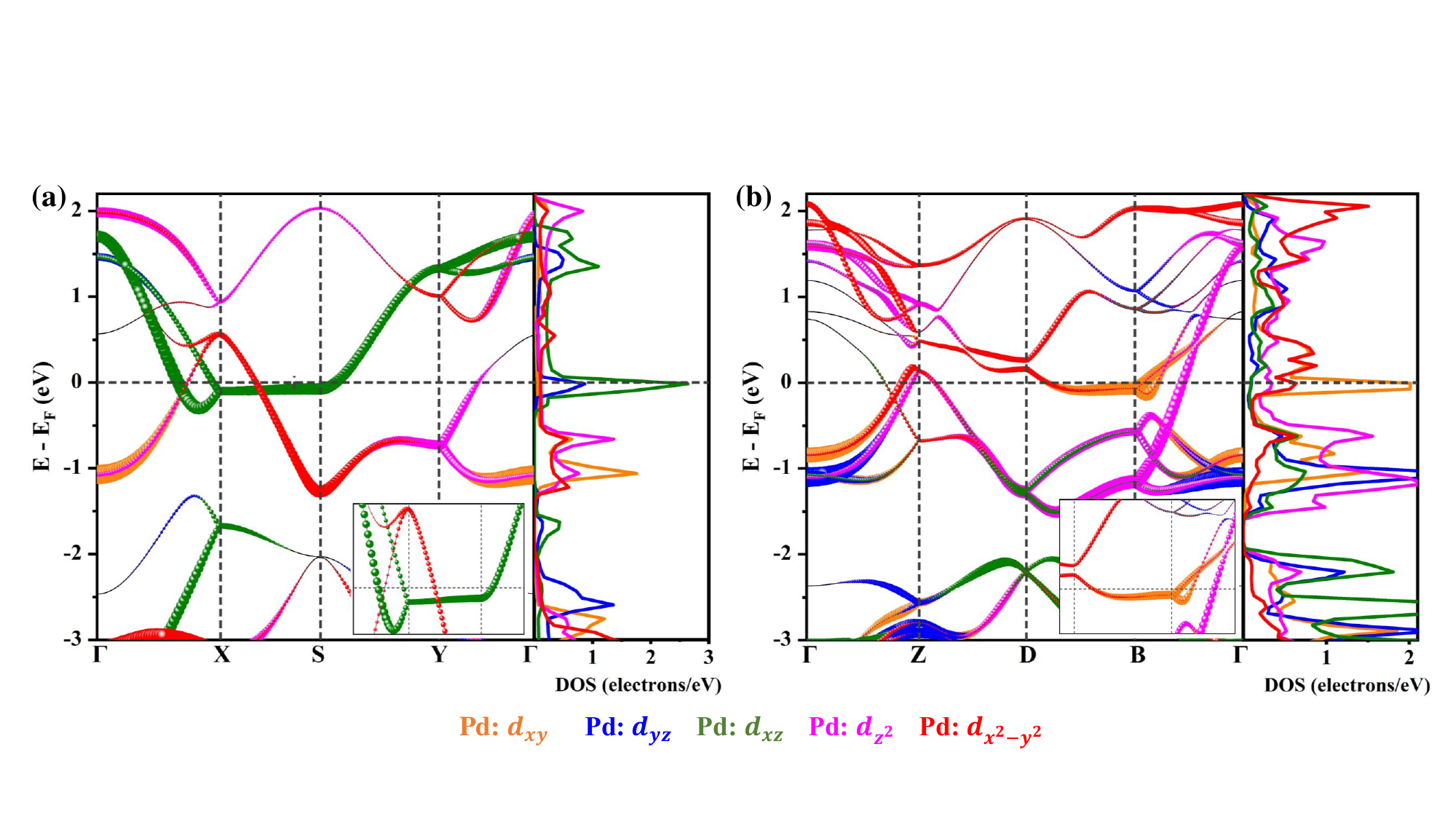}}
\caption{Orbital-projected band structure and density of states of (a) marcasite, and (b) arsenopyrite phases of bulk PdSe$_2$. Because of the octahedral distortion around 9 GPa, the energy of $d_{xz}$ orbital increases which contributes to the energy bands near Fermi level in case of marcasite phase. On the other hand, $d_{xy}$ orbital energetically uplifts in case of arsenopyrite phase. Both these orbitals in the new phases contribute flat bands across the Fermi level shown in a zoomed-in view at the inset of (a) and (b).     
\label{new phase band}}
\end{figure}

By analyzing the orbital-projected band structures, the electronic evolution driving the phase transition under high pressure were investigated. \fig{new phase band}(a) and \fig{new phase band}(b) represent the orbital-resolved band dispersion and density of states for the marcasite and arsenopyrite phases, respectively, which correspond to the contribution of the Pd $d$ orbitals to the formation of the energy bands near Fermi level. As discussed previously, pyrite PdSe$_2$ consists of symmetric octahedral coordination without any Jahn-Teller distortion and the equally populated $d_{x^2-y^2}$ and $d_{z^2}$ orbitals forming energy bands around the Fermi level, corresponds to a spin-triplet state ($t_{2g}^6 e_g^2$) as illustrated in Fig. S13 (middle panel) of SM \cite{supple}. Conversely, there is a predominant contribution of the $d_{xz}$ orbital for the emergent marcasite phase in forming the populated conduction band minima across the Fermi surface as depicted in \fig{new phase band}(a) and thereby preserves the persisting metallicity. The distortion appeared in the $t_{2g}$ orbitals corroborates with the octahedral compression along the $ac$ plane in the B-marcasite structure which breaks the existing degeneracy and energetically uplifts the $d_{xz}$ orbital (schematically shown in Fig. S13 right panel of SM \cite{supple}).
On the other hand, $d_{xy}$ orbital significantly dominates around the Fermi level for the another possible arsenopyrite structure as a result of the octahedral distortion generated within the $t_{2g}$ orbitals (Fig. S13 left panel of SM \cite{supple}).  
In contrast to pyrite TMDCs like MnS$_2$, MnTe$_2$ etc. \cite{MnS2-2,MnTe2}, where the pyrite to marcasite/arsenopyrite phase transition was governed by the high-spin ($t_{2g}^3 e_g^2$, S= 5/2) to low-spin ($t_{2g}^5 e_g^0$, S= 1/2) transition, PdSe$_2$ preserves the spin multiplicity in its high pressure marcasite/arsenopyrite phase. Accompanying the octahedral distortion in the newly stabilized high-pressure phases, the presence of flat bands primarily contributed by \(d_{xz}\) orbitals in the marcasite (\(Pnnm\)) phase and \(d_{xy}\) orbitals in the arsenopyrite (\(P2_1/c\)) phase near the Fermi level is anticipated to drive the emergence of strong-coupling superconductivity \cite{FB-1}. The formation of flat bands significantly reduces the Fermi velocity, \(v_F \propto \frac{\partial E}{\partial k}\), resulting in an infinitesimal coherence length, \(\xi \propto \frac{\hbar v_F}{\Delta}\), and a low superfluid stiffness, \(\rho_s \propto n_s v_F^2\), consistent with the framework of BCS theory. This, in turn, leads to the presence of massive, nearly immobile supercarriers that enhance the pairing strength \cite{FB-1,FB-2,FB-3,FB-4}.

Phonon band dispersions for all the HP phases are calculated using density functional perturbation theory (Fig. S14 of SM \cite{supple}). No imaginary mode is found at 12 GPa manifesting the stability of the HP phases at this pressure regime. Pressure-induced superconductivity in PdSe\(_2\), observed at \(T_c = 2.4\) K beyond 7 GPa, shows a rapid increase in \(T_c\) to 13.1 K under further compression up to 23 GPa, as reported by ElGhazali \textit{et al.} \cite{claudia}. This enhancement of \(T_c\) under pressure can be directly correlated with the emergence of flat, dispersionless bands near the Fermi level in the high-pressure marcasite and arsenopyrite phases. Conceptually, the proximity of these flat bands to the Fermi energy, as indicated by a sharp peak in the density of states \(N(E_F) \propto 1/\sqrt{E - E_F}\), amplifies the electron-phonon coupling through the McMillan relation, \(T_c \propto \exp(-1/\lambda)\), where \(\lambda \propto N(E_F) \langle g^2 \rangle / \omega^2\). The computationally estimated $T_c$ values for the pyrite, marcasite, and arsenopyrite phases of PdSe$_2$ at 12 GPa are 5.6 K, 6.8 K, and 5.8 K, respectively. This enhancement in $T_c$ facilitates robust Cooper pairing, further stabilizing the superconducting state. Such a mechanism highlights the critical role of band topology and orbital character in dictating superconducting properties under extreme conditions. 

\begin{figure}[ht!]
\centerline{\includegraphics[scale=0.55, clip]{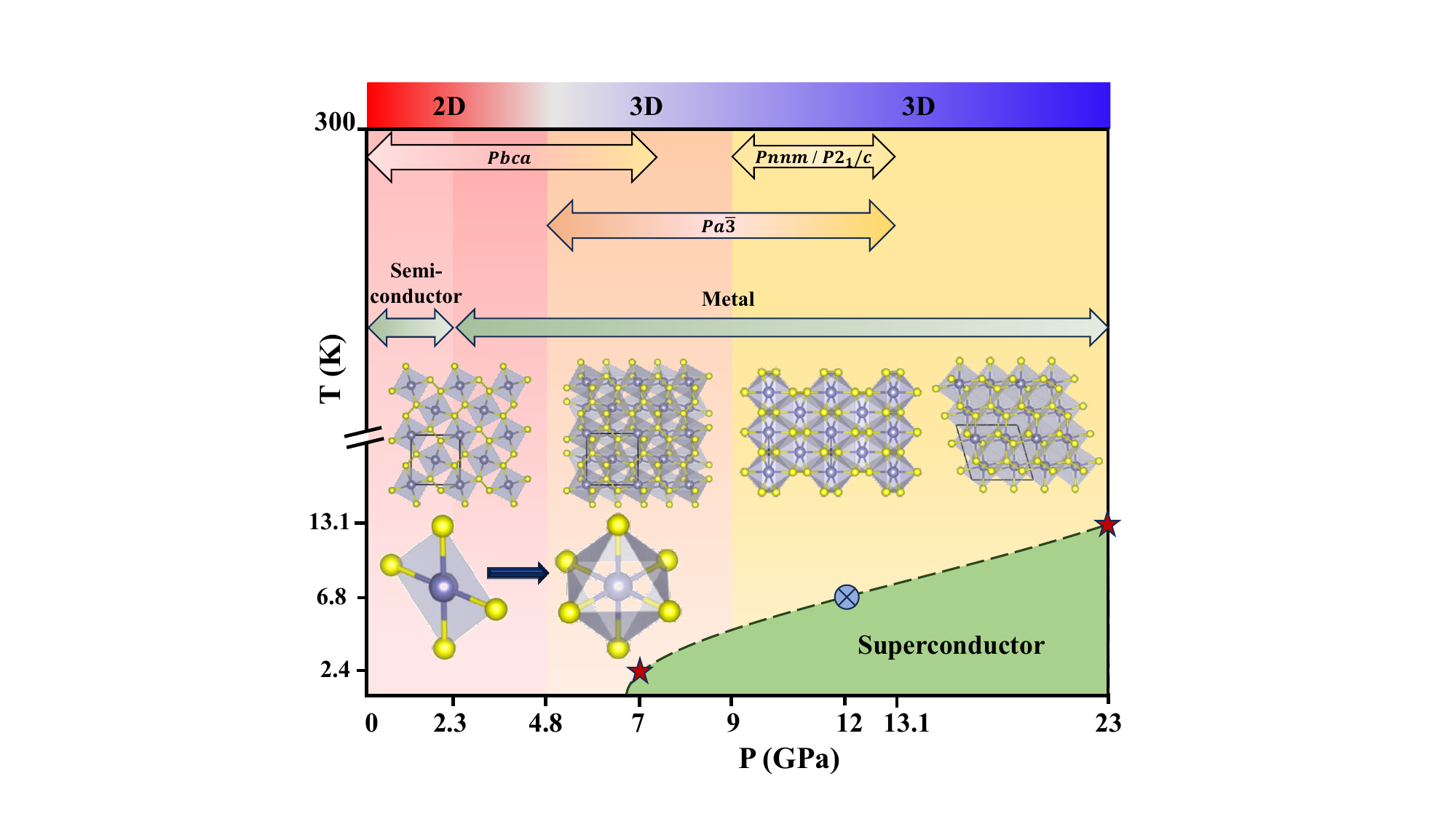}}
\caption{Schematic phase diagram encompassing all pressure-induced phases segmented according to the XRD-identified structural and electronic transitions. The superconducting domain commences from 7 GPa with $T_c$ = 2.4 K and is enhanced upto 13.1 K at 23 GPa as reported by Elghazali $et \ al.$ \cite{claudia} marked by asterisk symbols. The present study incorporates the structural phase transitions upto 13.1 GPa (experimentally accessible pressure regime) and calculates $T_c$ = 5.6-6.8 K at 12 GPa marked by circular symbol corresponding to the high pressure structures.
\label{phase diagram}}
\end{figure}

A comprehensive phase diagram encompassing all pressure-induced phases alongside the superconducting domain schematically depicted in \fig{phase diagram}, serving to encapsulate the entirety of this study in a single illustration. The regimes are segmented according to the XRD-identified structural and electronic transitions. 

\section{Conclusions}
\label{sec:conclusion} 
In conclusion, we have explored the pressure-induced structural and electronic transitions in PdSe$_2$ using high-pressure X-ray diffraction, Raman spectroscopy, and first-principles calculations. At pressure above 2.3 GPa, the suppression of the Jahn-Teller distortion drives lattice expansion and metallization, culminating in a 2D to 3D dimensional crossover at 4.8 GPa. This transition is marked by interlayer contraction and charge redistribution, stabilizing the pyrite phase. At higher pressures ($\sim$ 9 GPa), a novel phase emerges with octahedral distortions in Pd $t_{2g}$ orbitals. Structural refinements indicate the stabilization of either the marcasite (\textit{Pnnm}) or arsenopyrite (\textit{P2$_1$/c}) phase under high-pressure conditions. Raman spectroscopy and band structure analyses reveal semiconducting-to-metallic transitions, spin-state modifications, and flat-band characteristics, providing insight into the origins of pressure-induced strong-coupling superconductivity. These results highlight the tunable structural and electronic properties of PdSe$_2$, emphasizing its role as a promising material for exploring pressure-driven phase transitions.


\begin{acknowledgments}
IACS, DST-INSPIRE, and UGC are greatly acknowledged for fellowships. T.K. would like to thank XPRESS beamline, Elettra Sincrotrone, Italy. Financial support through Italian Ministry of Foreign Affairs and International Cooperation and the Indian Department of Science and Technology is greatly acknowledged. The authors are thankful to Achintya Singha, Sandip Dhara, Sanjay Kumar, and Anand Kumar for their technical help during the high-pressure Raman experiments. T.K. also acknowledges IACS cluster facility. T.K. is thankful to Supriyo Santra for useful discussion. S. Datta acknowledges the financial support from DST-ANRF under grant No. CRG/2021/004334 and ECR/2017/002037 and UGC-DAE CSR grant No. CRS-M-279. S. Datta also acknowledges support from the Technical Research Centre (TRC), IACS, Kolkata.

S. Datta conceived the project. T.K. performed the sample growth and characterizations. T.K., B.J., and S.K.M. performed the synchrotron XRD. T.K., and K. Dey analyzed the high pressure XRD data. M.P., and S. Das designed the high-pressure experimental setup for Raman spectroscopy. S. Das, T.K., and C.N. performed the Raman measurements and analyzed the data. K. Dolui, and T.K. performed the theoretical calculations. All authors discussed the results and actively commented on the manuscript written by T.K..

\end{acknowledgments}

\end {document}


\title{Supporting Information: 

Dimensional Crossover and Emergence of Novel Phases in Puckered PdSe$_2$
under Pressure}

\author{Tanima Kundu}
\email{tanima.kundu96@gmail.com}
\affiliation{School of Physical Sciences, Indian Association for the Cultivation of Science, 2A $\&$ B
Raja S. C. Mullick Road, Jadavpur, Kolkata - 700032, India
}

\author{Soumik Das}
\affiliation{School of Physical Sciences, Indian Association for the Cultivation of Science, 2A $\&$ B
Raja S. C. Mullick Road, Jadavpur, Kolkata - 700032, India
}

\author{Koushik Dey}
\affiliation{School of Physical Sciences, Indian Association for the Cultivation of Science, 2A $\&$ B
Raja S. C. Mullick Road, Jadavpur, Kolkata - 700032, India
}

\author{Boby Joseph}
\affiliation{Elettra - Sincrotrone Trieste S.C. p. A., S.S. 14, Km 163.5 in Area Science Park, Basovizza 34149, Italy}

\author{Chumki Nayak}
\affiliation{Department of Physical Sciences, Bose Institute, Unified academic campus, EN 80, Sector V, Bidhan Nagar, Kolkata- 700091, India}

\author{Mainak Palit}
\affiliation{School of Physical Sciences, Indian Association for the Cultivation of Science, 2A $\&$ B
Raja S. C. Mullick Road, Jadavpur, Kolkata - 700032, India
}

\author{Sanjoy Kr Mahatha} 
\affiliation{UGC-DAE Consortium for Scientific Research, Khandwa Road, Indore 452001, Madhya Pradesh, India}

\author{Kapildeb Dolui}
\affiliation{Department of Materials Science \& Metallurgy, University of Cambridge, 27 Charles Babbage Road, Cambridge CB3 0FS, United Kingdom}
\affiliation{Department of Physics, Indian Institute of Technology Tirupati, Tirupati, Andhra Pradesh 517619, India}

\author{Subhadeep Datta}
\email{sspsdd@iacs.res.in}
\affiliation{School of Physical Sciences, Indian Association for the Cultivation of Science, 2A $\&$ B
Raja S. C. Mullick Road, Jadavpur, Kolkata - 700032, India
}

\maketitle

\begin{table}[h]
    \centering
    \small
    \setlength{\tabcolsep}{3pt} 
    \renewcommand{\arraystretch}{1}
    \renewcommand{\thetable}{\Roman{table}}   
    
    \begin{tabular}{|c|c|c|c|c|c|c|c|c|c|}
        \hline
        Space & P (GPa) & Lattice Parameters (\AA) & Atoms & Wyckoff & $x$ & $y$ & $z$ & R$_p$ & R$_{wp}$ \\
        Group & & & & & & & & & \\
        \hline
         & Ambient & a = 5.7452(2), b = 5.8682(3), & Pd & 4a & 0.0000 & 0.0000 & 0.0000 & 14.5 & 11.3 \\
         & & c = 7.6974(9) & Se & 8c & 0.3922 & 0.3856 & 0.4021 & & \\
         \cline{2-10}
         
         $Pbca$& 2.3 & a = 5.7341(8), b = 5.8439(7), & Pd & 4a & 0.0000 & 0.0000 & 0.0000 & 13.8 & 9.9 \\
         & & c = 7.3318(7) & Se & 8c & 0.3855 & 0.3682 & 0.3916 & & \\
         \cline{2-10}         
         
        & 4.8 & a = 5.8654(3), b = 5.8882(3), & Pd & 4a & 0.0000 & 0.0000 & 0.0000 & 10.6 & 8.5 \\
        & & c = 6.7128(3) & Se & 8c & 0.3744 & 0.3903 & 0.3822 & & \\
        \hline
         & 4.8 & a = b = c = 6.0892(6) & Pd & 4a & 0.0000 & 0.0000 & 0.0000 & 10.6 & 8.5 \\
         & & & Se & 8c & 0.3901 & 0.3901 & 0.3901 & & \\
         \cline{2-10}         
         
        & 7.2 & a = b = c = 6.0504(5) & Pd & 4a & 0.0000 & 0.0000 & 0.0000 & 10.0 & 8.5 \\
        & & & Se & 8c & 0.3892 & 0.3892 & 0.3892 & & \\
        \cline{2-10}        
        
        $Pa\bar{3}$ & 8.4 & a = b = c = 6.0334(6) & Pd & 4a & 0.0000 & 0.0000 & 0.0000 & 9.6 & 7.6 \\
        & & & Se & 8c & 0.3883 & 0.3883 & 0.3883 & & \\
        \cline{2-10}
        
        & 9.2 & a = b = c = 6.0406(3) & Pd & 4a & 0.0000 & 0.0000 & 0.0000 & 9.7 & 8.6 \\
        & & & Se & 8c & 0.3860 & 0.3860 & 0.3860 & & \\
        \cline{2-10}
        
        & 11.1 & a = b = c = 6.0206(7) & Pd & 4a & 0.0000 & 0.0000 & 0.0000 & 9.0 & 7.8 \\
        & & & Se & 8c & 0.3837 & 0.3837 & 0.3837 & & \\
        \cline{2-10}
        
        & 13.1 & a = b = c = 5.9959(7) & Pd & 4a & 0.0000 & 0.0000 & 0.0000 & 15.5 & 11.7 \\
        & & & Se & 8c & 0.3834 & 0.3834 & 0.3834 & & \\
        \hline
    \end{tabular}
    \caption{Table of structural parameters obtained from Rietveld refinement.}
    \label{tab:rietveld}
\end{table}

\begin{table}[h]
    \centering
    \small  
    \setlength{\tabcolsep}{2pt} 
    \renewcommand{\arraystretch}{1} 
    \renewcommand{\thetable}{\Roman{table}} 
    
    \begin{tabular}{|c|c|c|c|c|c|c|c|c|c|}
        \hline
        Space & P (GPa) & Lattice Parameters (\AA) & Atoms & Wyckoff & $x$ & $y$ & $z$ & R$_p$ & R$_{wp}$ \\
        Group & & & & & & & & & \\
        \hline
         & 9.7 & a = 4.6942(7), b = 5.1945(4), & Pd & 2a & 0.0000 & 0.0000 & 0.0000 & 10.5 & 9.6 \\
         & & c = 3.9993(3) & Se & 4g & 0.2217 & 0.5987 & 0.0000 & & \\
         \cline{2-10}
        
        & 10.2 & a = 4.6989(9), b = 5.1741(8), & Pd & 2a & 0.0000 & 0.0000 & 0.0000 & 11.1 & 10.5 \\
         & & c = 3.9931(5) & Se & 4g & 0.2221 & 0.5978 & 0.0000 & & \\      
        \cline{2-10}
         
         $Pnnm$ & 10.6 & a = 4.7088(7), b = 5.1432(5), & Pd & 2a & 0.0000 & 0.0000 & 0.0000 & 10.6 & 9.1 \\
         & & c = 3.9926(4) & Se & 4g & 0.7772 & 0.5958 & 0.0000 & & \\
         \cline{2-10}         
         
        & 11.1 & a = 4.7205(7), b = 5.0861(5), & Pd & 2a & 0.0000 & 0.0000 & 0.0000 & 9.0 & 7.8 \\
        & & c = 3.9776(5) & Se & 4g & 0.7765 & 0.5934 & 0.0000 & & \\
        \cline{2-10}
        & 13.1 & a = 4.7298(2), b = 5.0419(7), & Pd & 2a & 0.0000 & 0.0000 & 0.0000 & 15.5 & 11.7 \\
        & & c = 3.9565(9) & Se & 4g & 0.7748 & 0.5889 & 0.0000 & & \\        
        \hline
         
         & 9.7 & a = 6.0883(3), b = 5.9866(3), & Pd & 4e & 0.7676 & 0.4364 & 0.7885 & 10.1 & 8.0 \\
        & & c = 6.0524(5) & Se & 4e & 0.7776 & 0.8436 & 0.6276 & & \\
         & & & Se & 4e & 0.3366 & 0.6337 & 0.6407 & & \\
         \cline{2-10}         
         
        & 10.2 & a = 6.0923(2), b = 5.9781(3), & Pd & 4e & 0.8196 & 0.5184 & 0.6957 & 9.8 & 8.1 \\
        & & c = 6.0356(4) & Se & 4e & 0.8198 & 0.8318 & 0.6747 & & \\
        & & & Se & 4e & 0.4026 & 0.6396 & 0.7378 & & \\
        \cline{2-10}
         $P2_1/c$ & 10.6 & a = 6.0943(3), b = 5.9728(4), & Pd & 4e & 0.7517 & 0.4978 & 0.7445 & 9.7 & 8.2 \\
         & & c = 6.0134(4) & Se & 4e & 0.8367 & 0.8838 & 0.6407 & & \\
         & & & Se & 4e & 0.3412 & 0.6145 & 0.6563 & & \\
         \cline{2-10}         
         
        & 11.1 & a = 6.0901(3), b = 5.9571(4), & Pd & 4e & 0.7905 & 0.5455 & 0.7576 & 8.7 & 7.4 \\
        & & c = 5.9977(6) & Se & 4e & 0.7829 & 0.8689 & 0.6367 & & \\
        & & & Se & 4e & 0.3529 & 0.6327 & 0.6931 & & \\
        \cline{2-10}
        & 13.1 & a = 6.0512(3), b = 5.9087(1), & Pd & 4e & 0.8191 & 0.5171 & 0.7091 & 9.1 & 7.2 \\
        & & c = 5.9727(6) & Se & 4e & 0.8046 & 0.8426 & 0.6299 & & \\
        & & & Se & 4e & 0.3654 & 0.6316 & 0.7031 & & \\
        \hline
        
    \end{tabular}
    \caption{Table of structural parameters obtained from Rietveld refinement.}
    \label{tab:HP-phase}
\end{table}

\begin{table}[h]
    \centering
    \small
    \setlength{\tabcolsep}{35pt} 
    \renewcommand{\arraystretch}{0.5}
    \renewcommand{\thetable}{\Roman{table}}  
    
    \begin{tabular}{|c|c|c|}
        \hline
        \multirow{2}{4em}{P (GPa)} 
        & \multicolumn{2}{|c|}{Phase Fraction (\%)} \\
        \cline{2-3}
         & $Pbca$ & $Pa\bar{3}$  \\
         \cline{2-3}
         \hline         
         Ambient & 100 &  \\
         \cline{1-2}         
        
         0.5 & 100 &  \\
         \cline{1-2}
                  
         1.0 & 100 &  \\
         \cline{1-2}
                  
         2.3 & 100 &  \\
         \cline{1-2}
                  
         3.5 & 100 &  \\
         \cline{1-2}
                  
         4.4 & 100 &  \\
         \cline{1-3}
                  
         4.8 & 30 & 70  \\
         \cline{1-3}
         
         5.3 & 5.64 & 94.36 \\
         \cline{1-3}
          
         5.8 & 1.67 & 98.33 \\
         \cline{1-3}
         
         6.3 & 1.62 & 98.38 \\
         \cline{1-3}
          
         6.7 & 0.76 & 99.24 \\
         \cline{1-3}
          
         7.2 &  & 100 \\
         \cline{1-1}
         \cline{3-3}
          
         7.6 &  & 100  \\
         \cline{1-1}
         \cline{3-3}
        
         8.0 &  & 100  \\
         \cline{1-1}
         \cline{3-3}
         
         8.4 &  & 100  \\
         \cline{1-3}
                
    \end{tabular}
    \caption{Table of phase fraction of $Pbca$ and $Pa\bar{3}$ phases obtained from Rietveld refinement.}
    \label{tab:rietveld}
\end{table}

\begin{table}[h]
    \centering
    \small
    \setlength{\tabcolsep}{35pt} 
    \renewcommand{\arraystretch}{0.5}
    \renewcommand{\thetable}{\Roman{table}}  
    
    \begin{tabular}{|c|c|c|}
        \hline
        \multirow{2}{4em}{P (GPa)} 
        & \multicolumn{2}{|c|}{Phase Fraction (\%)} \\
        \cline{2-3}
         & $Pa\bar{3}$ & $Pnnm$  \\
         \cline{1-3}
                  
         8.8 & 100 & 0  \\
         \cline{1-3}         
        
         9.2 & 94.01 & 5.99  \\
         \cline{1-3}
                
         9.7 & 85.50 & 14.50  \\
         \cline{1-3}
                
         10.2 & 84.55 & 15.45 \\
         \cline{1-3}
                
         10.6 & 84.03 & 15.97  \\
         \cline{1-3}
                
         11.1 & 82.42 & 17.58 \\
         \cline{1-3}
                
         13.1 & 78.04 & 21.96  \\
         \cline{1-3}
          
    \end{tabular}
    \caption{Table of phase fraction of $Pa\bar{3}$ and $Pnnm$ phases obtained from Rietveld refinement.}
    \label{tab:rietveld}
\end{table}

\begin{table}[h]
    \centering
    \small
    \setlength{\tabcolsep}{35pt} 
    \renewcommand{\arraystretch}{0.5} 
    \renewcommand{\thetable}{\Roman{table}} 
    
    \begin{tabular}{|c|c|c|}
        \hline
        \multirow{2}{4em}{P (GPa)} 
        & \multicolumn{2}{|c|}{Phase Fraction (\%)} \\
        \cline{2-3}
         & $Pa\bar{3}$ & $P2_1/c$  \\
         \cline{1-3}
                
         8.8 & 88.66 & 11.34  \\
         \cline{1-3}         
        
         9.2 & 81.75 & 18.25  \\
         \cline{1-3}
                
         9.7 & 71.80 & 28.20  \\
         \cline{1-3}
                
         10.2 & 67.63 & 32.37 \\
         \cline{1-3}
                
         10.6 & 66.70 & 33.30  \\
         \cline{1-3}
                 
         11.1 & 62.72 & 37.28 \\
         \cline{1-3}
                  
         13.1 & 58.10 & 41.90  \\
         \cline{1-3}
                 
    \end{tabular}
    \caption{Table of phase fraction of $Pa\bar{3}$ and $P2_1/c$ phases obtained from Rietveld refinement.}
    \label{tab:rietveld}
\end{table}

\clearpage

\begin{figure}[ht!]
\centerline{\includegraphics[scale=0.55, clip]{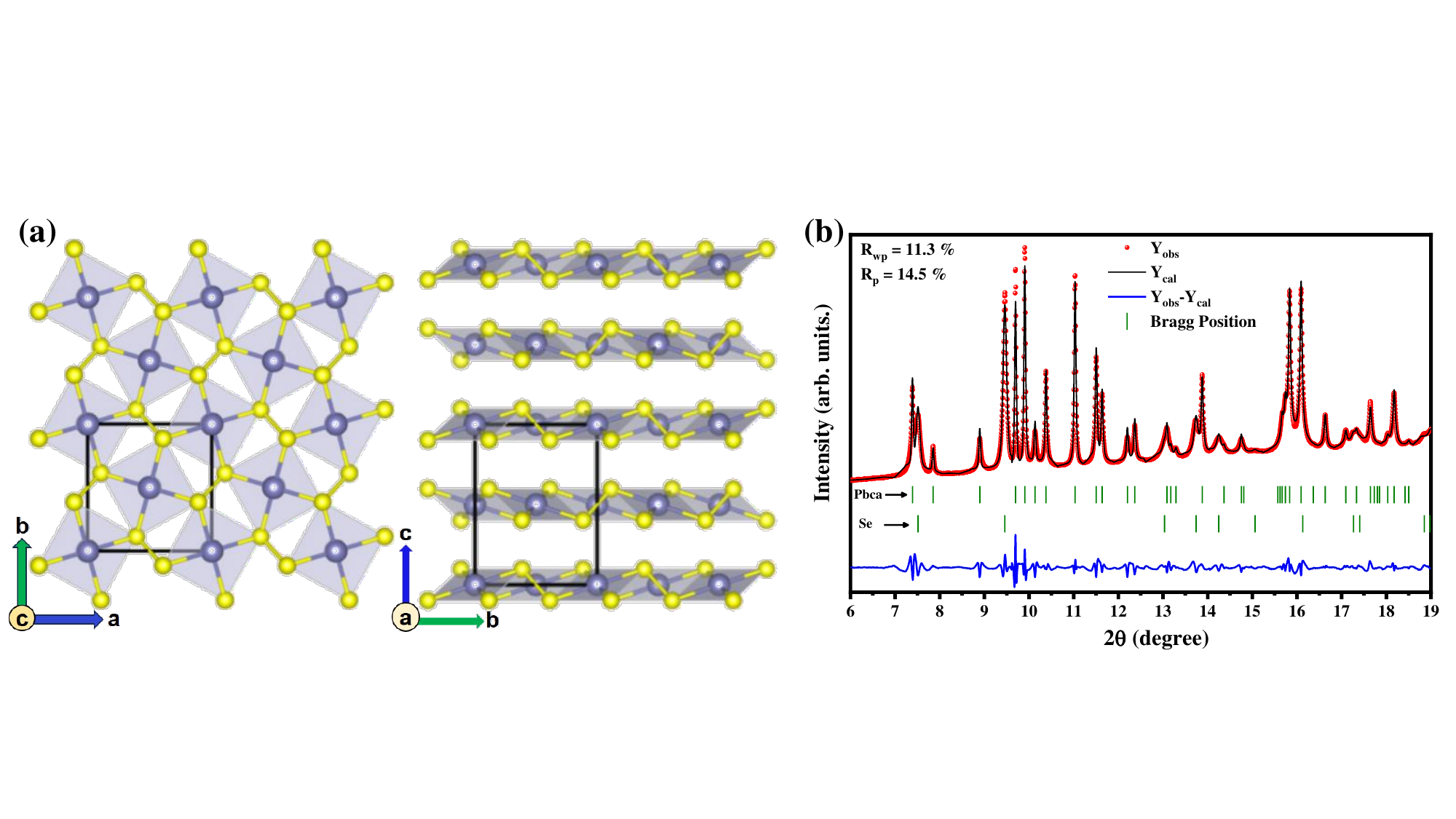}}
\renewcommand{\thefigure}{S\arabic{figure}}
\caption{Ambient PdSe$_2$: (a) Top and side view of the orthorhombic $Pbca$ crystal structure revealing the puckered pentagonal vdW nature of ambient PdSe$_2$. (b) Rietveld refined synchrotron powder X-ray diffraction pattern at ambient. The Bragg peaks are well-indexed with orthorhombic $Pbca$ phase mixed with a few percentage of Se flux.
\label{ambient}}
\end{figure}

\begin{figure}[ht!]
\centerline{\includegraphics[scale=0.55, clip]{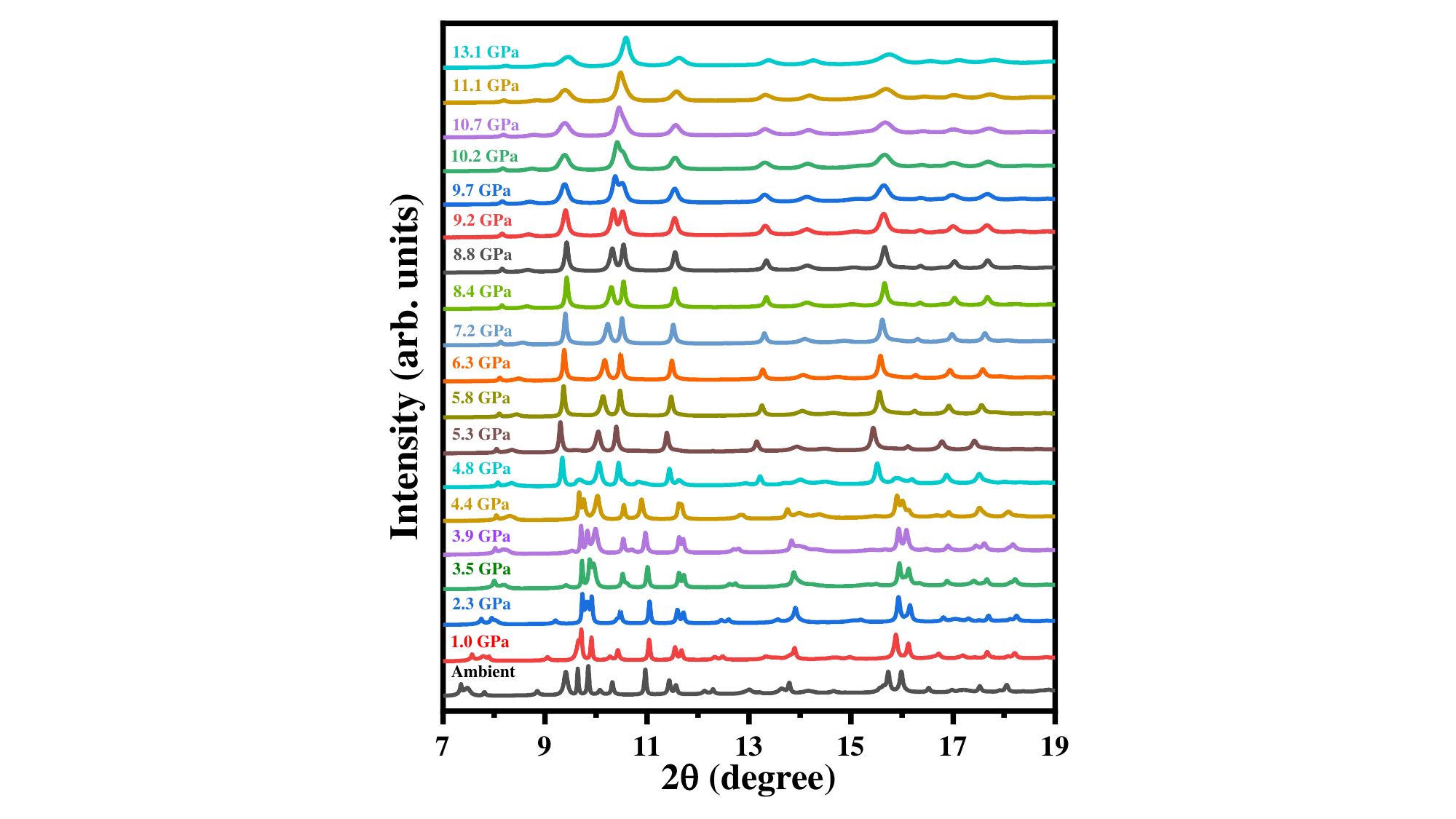}}
\renewcommand{\thefigure}{S\arabic{figure}}
\caption{Stack plot containing the synchrotron powder XRD pattern at different pressures from ambient to 13.1 GPa.
\label{stack plot}}
\end{figure}

\begin{figure}[ht!]
\centerline{\includegraphics[scale=0.55, clip]{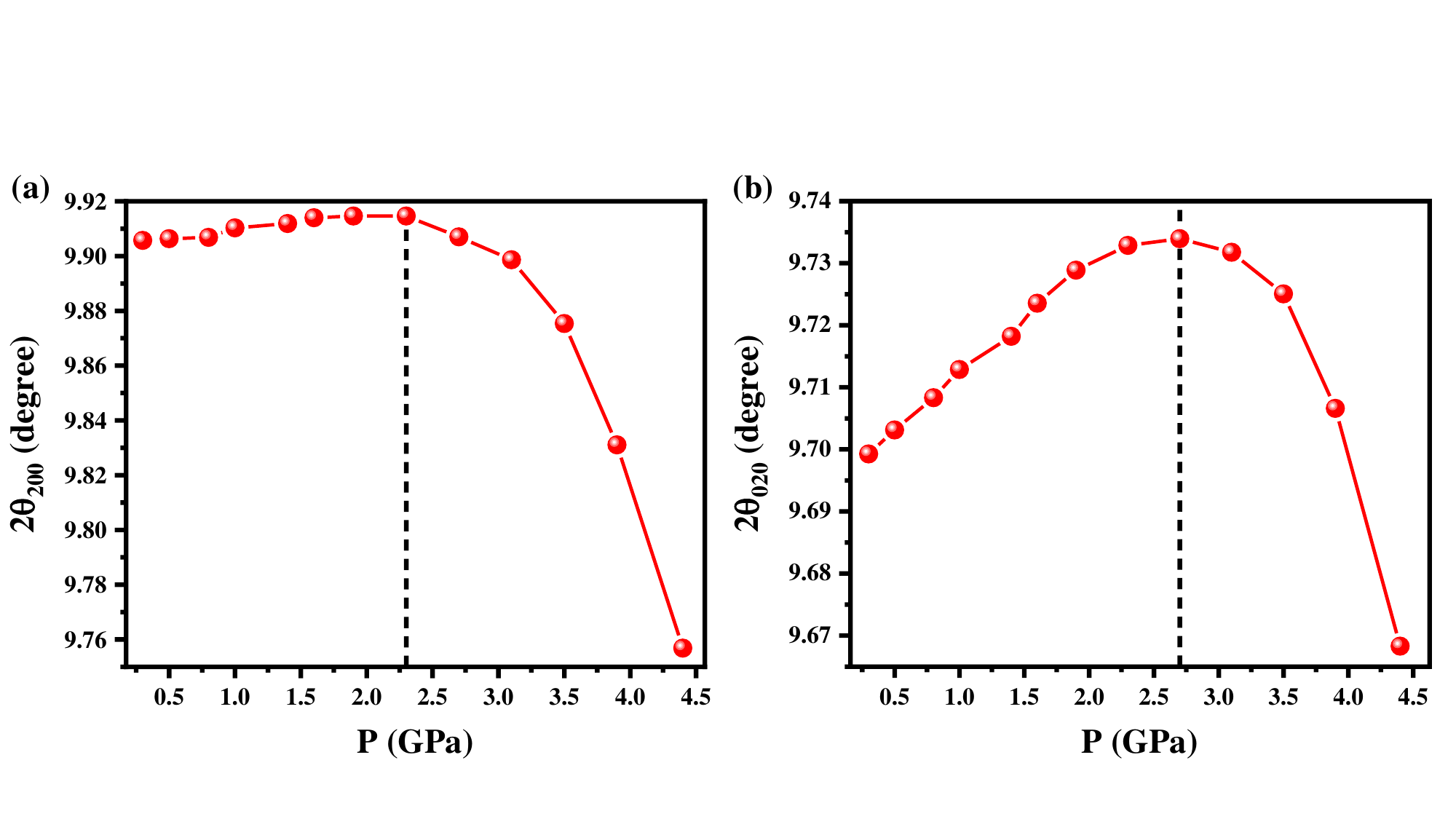}}
\renewcommand{\thefigure}{S\arabic{figure}}
\caption{2$\theta$ $vs.$ pressure plot for (a) (200) and, (b) (020) Bragg peaks showing anomalous shifting of in-plane Bragg peaks towards lower 2$\theta$ beyond 2.3-2.7 GPa.
\label{theta}}
\end{figure}

\begin{figure}[ht!]
\centerline{\includegraphics[scale=0.55, clip]{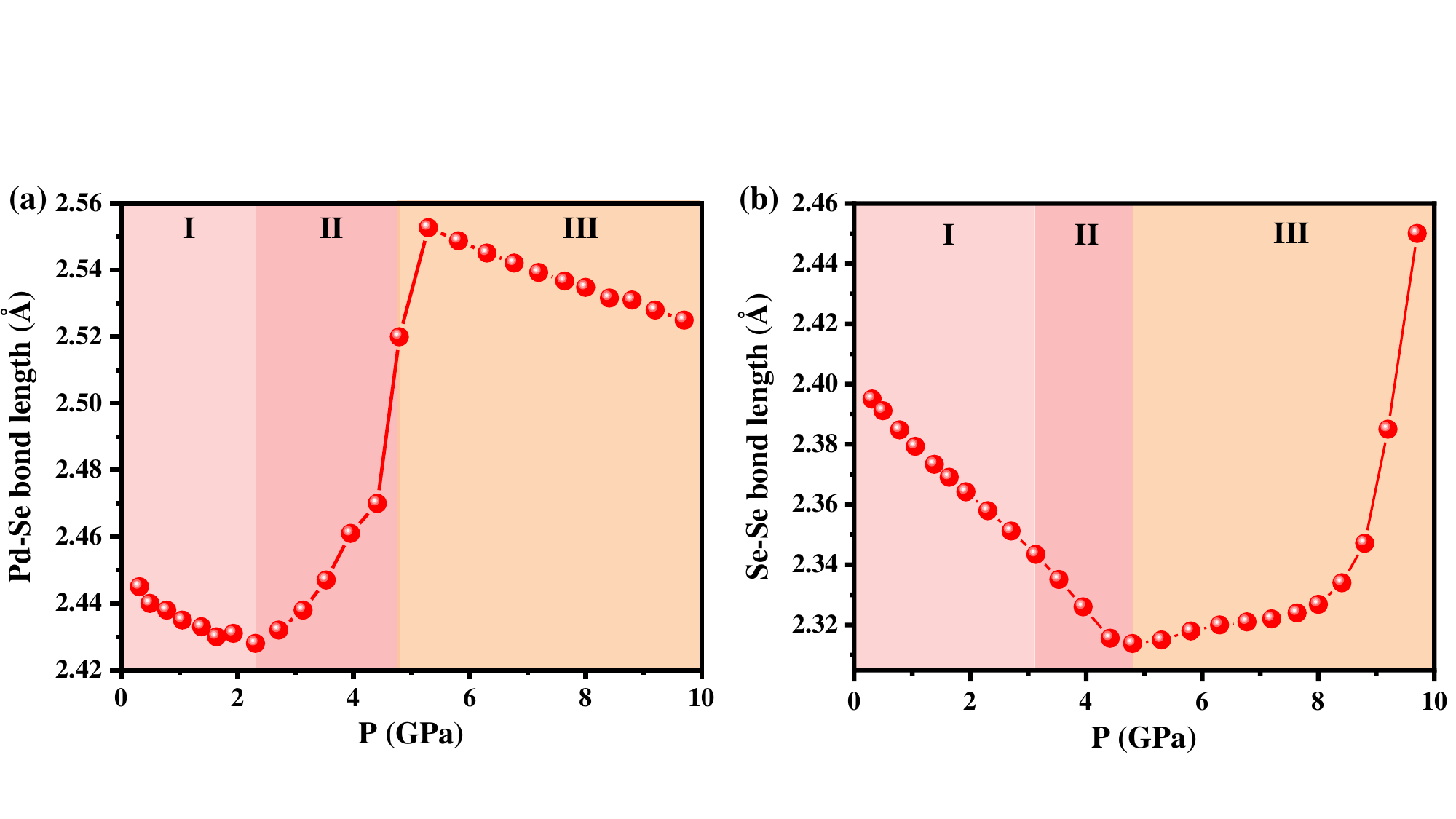}}
\renewcommand{\thefigure}{S\arabic{figure}}
\caption{(a) Pd$-$Se bond lengths are plotted against pressure. The regime I indicates the usual decrease in bond lengths under pressure manifesting the lattice contraction. At regime II, increase in the Pd$-$Se bond length is associated with the intralayer charge distribution and in-plane lattice expansion. Regime III resembles the pyrite phase with the usual decrease in Pd$-$Se bond length. (b) Se$-$Se bond length decreases under pressure in regime I and II, then keeps increasing throughout the existence of pyrite phase.    
\label{bond length}}
\end{figure}

\begin{figure}[ht!]
\centerline{\includegraphics[scale=0.65, clip]{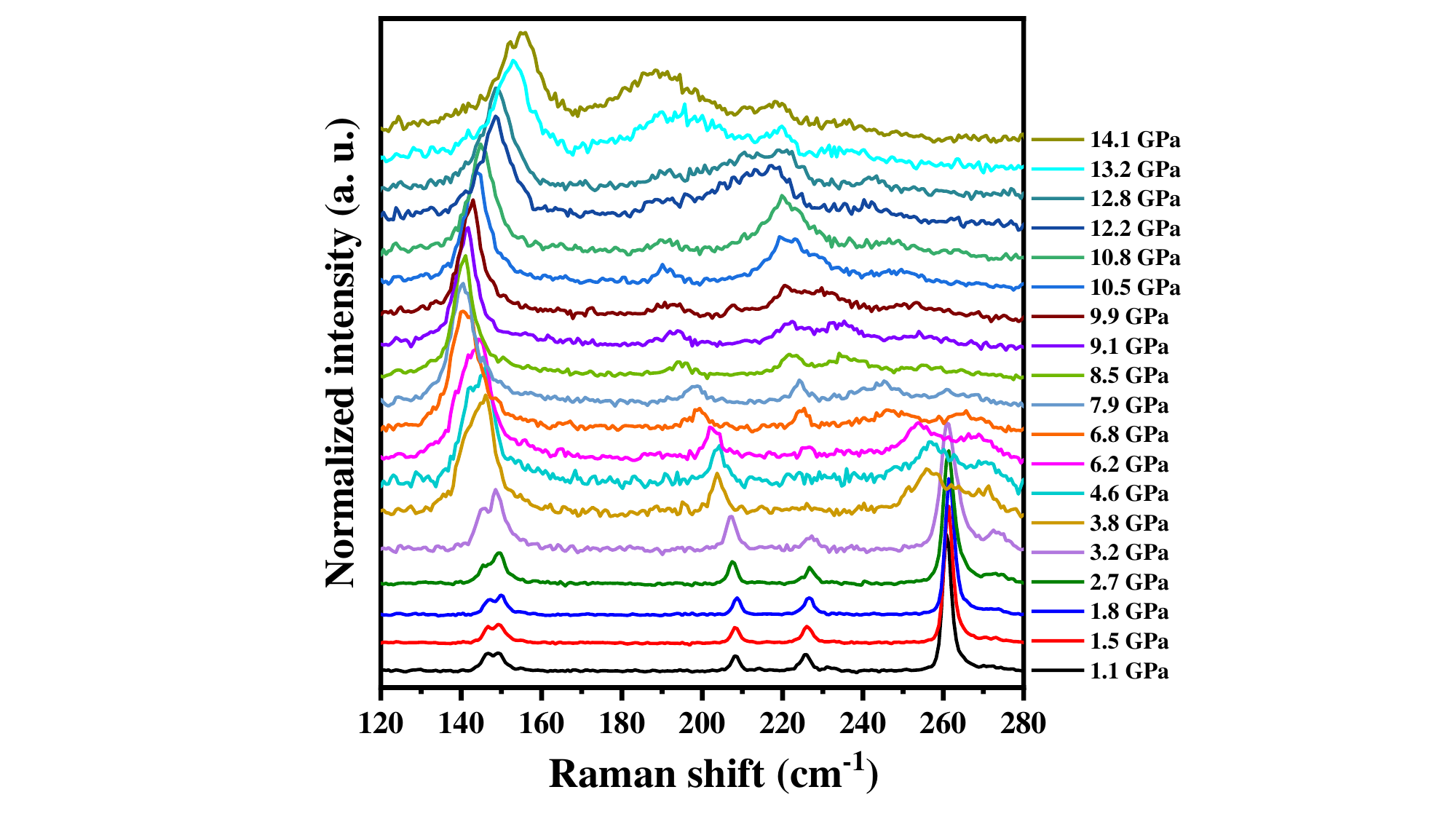}}
\renewcommand{\thefigure}{S\arabic{figure}}
\caption{Stacked Raman spectra of PdSe$_2$ under different pressure conditions measured at room temperature.
\label{stack Raman}}
\end{figure}

\begin{figure}[ht!]
\centerline{\includegraphics[scale=0.6, clip]{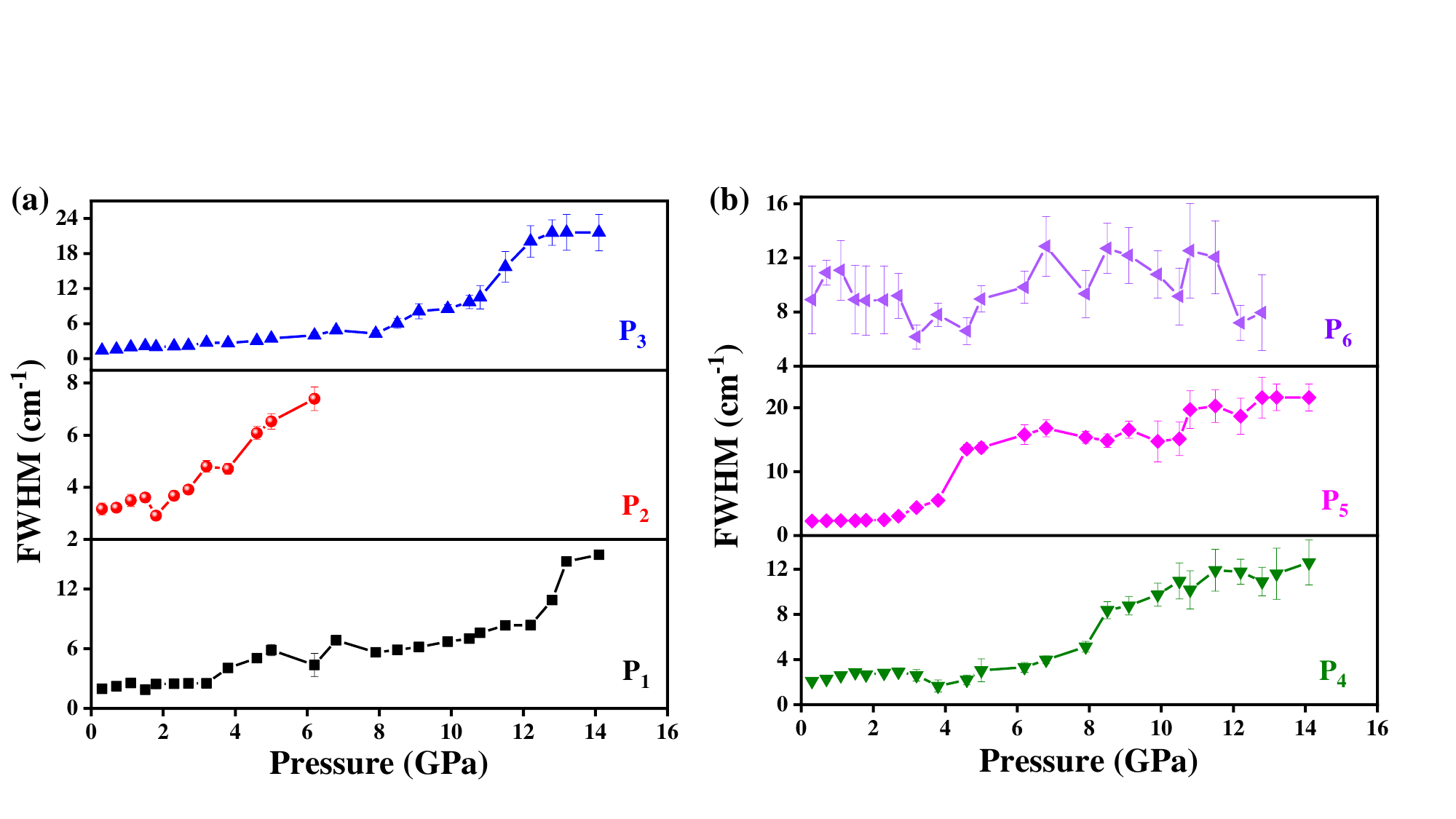}}
\renewcommand{\thefigure}{S\arabic{figure}}
\caption{(a) FWHM vs. pressure plots for (a) P1, P2, and P3 modes, and (b) P4, P5, and P6 modes.
\label{fwhm}}
\end{figure}

\begin{figure}[ht!]
\centerline{\includegraphics[scale=0.65, clip]{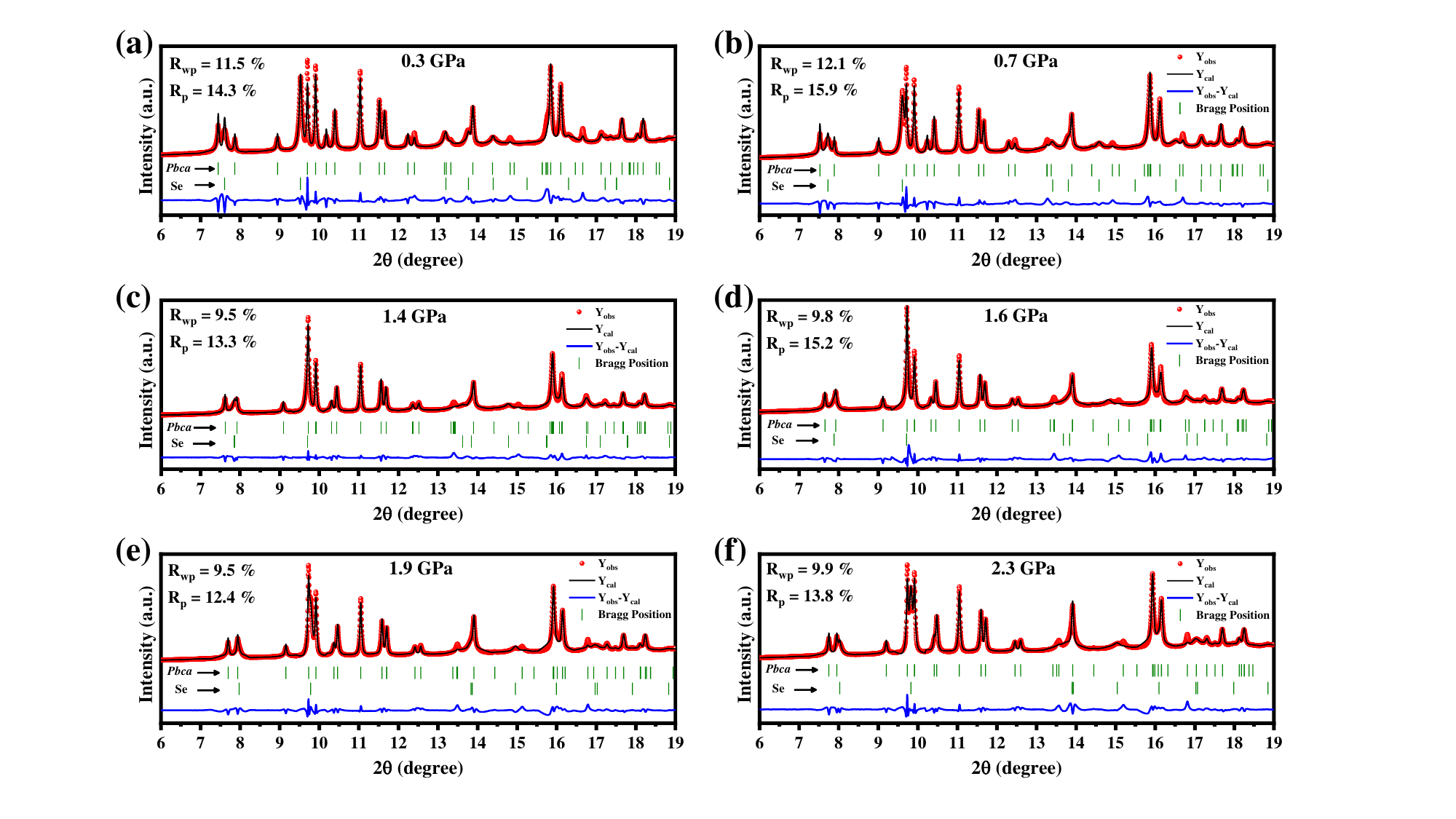}}
\renewcommand{\thefigure}{S\arabic{figure}}
\caption{(a) Rietveld refined XRD pattern at (a) 0.3 GPa, (b) 0.7 GPa, (c) 1.4 GPa, (d) 1.6 GPa, (e) 1.9 GPa, and (f) 2.3 GPa, corresponding to the ambient $Pbca$ phase of PdSe$_2$ mixed with Se-flux. 
\label{refined1}}
\end{figure}

\begin{figure}[ht!]
\centerline{\includegraphics[scale=0.65, clip]{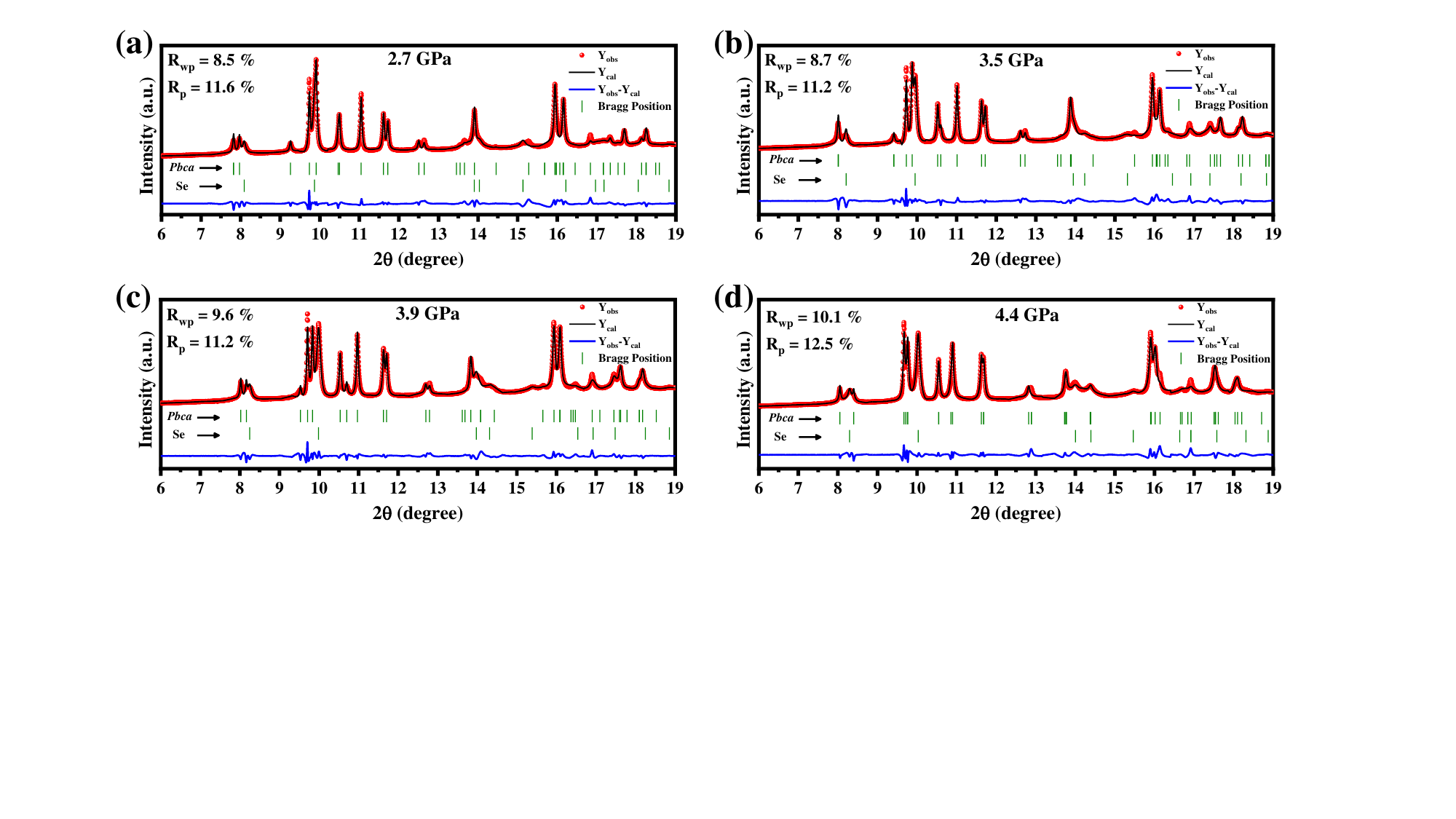}}
\renewcommand{\thefigure}{S\arabic{figure}}
\caption{(a) Rietveld refined XRD pattern at (a) 2.7 GPa, (b) 3.5 Pa, (c) 3.9 GPa, and (d) 4.4 GPa, corresponding to the ambient $Pbca$ phase of PdSe$_2$ mixed with Se-flux.
\label{refined2}}
\end{figure}

\begin{figure}[ht!]
\centerline{\includegraphics[scale=0.65, clip]{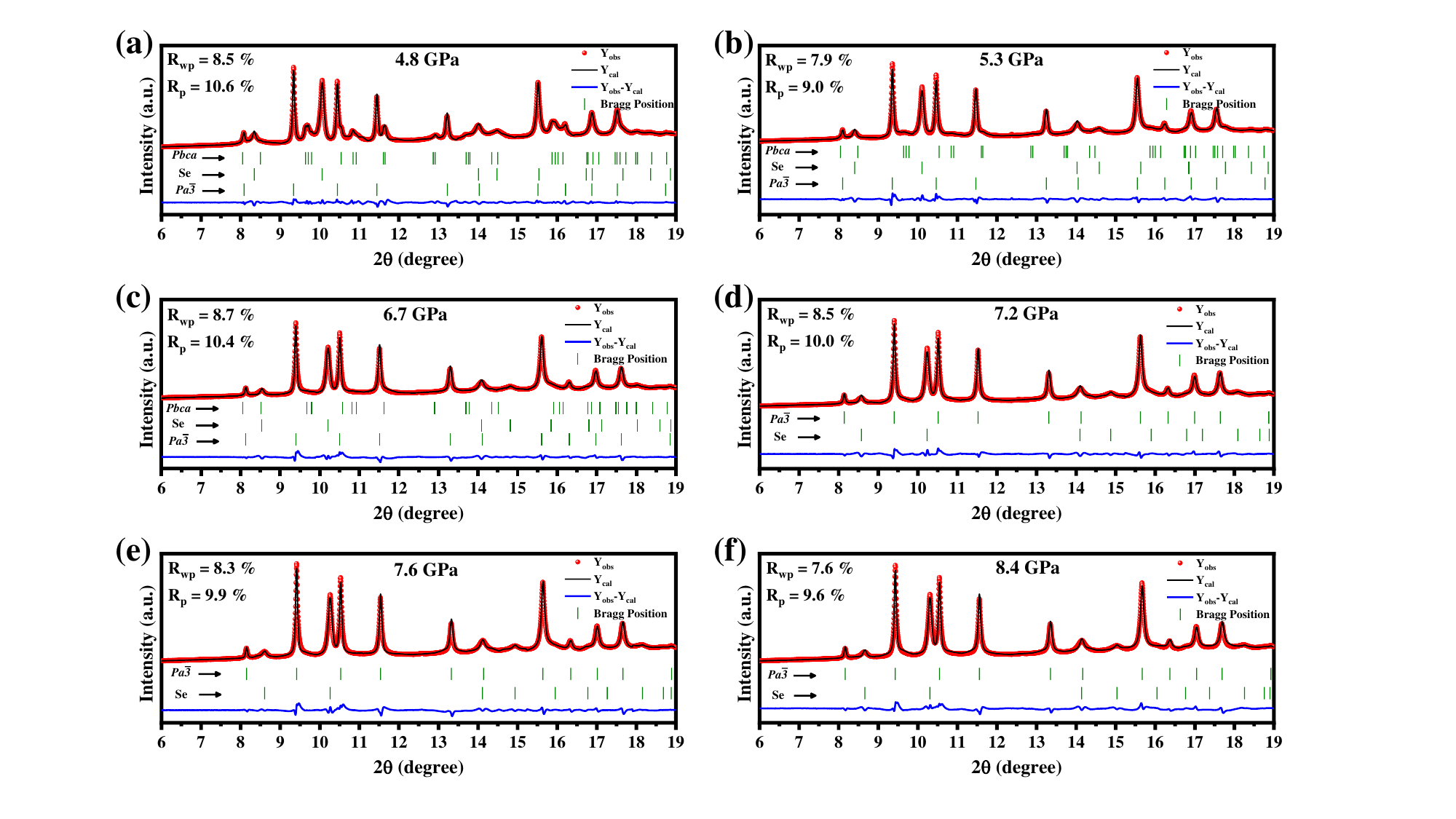}}
\renewcommand{\thefigure}{S\arabic{figure}}
\caption{(a) Rietveld refined XRD pattern at (a) 4.8 GPa, (b) 5.3 GPa, and (c) 6.7 GPa, describing the phase coexistence of ambient ($Pbca$) and pyrite ($Pa\bar{3}$) phases.
\label{refined3}}
\end{figure}

\begin{figure}[ht!]
\centerline{\includegraphics[scale=0.65, clip]{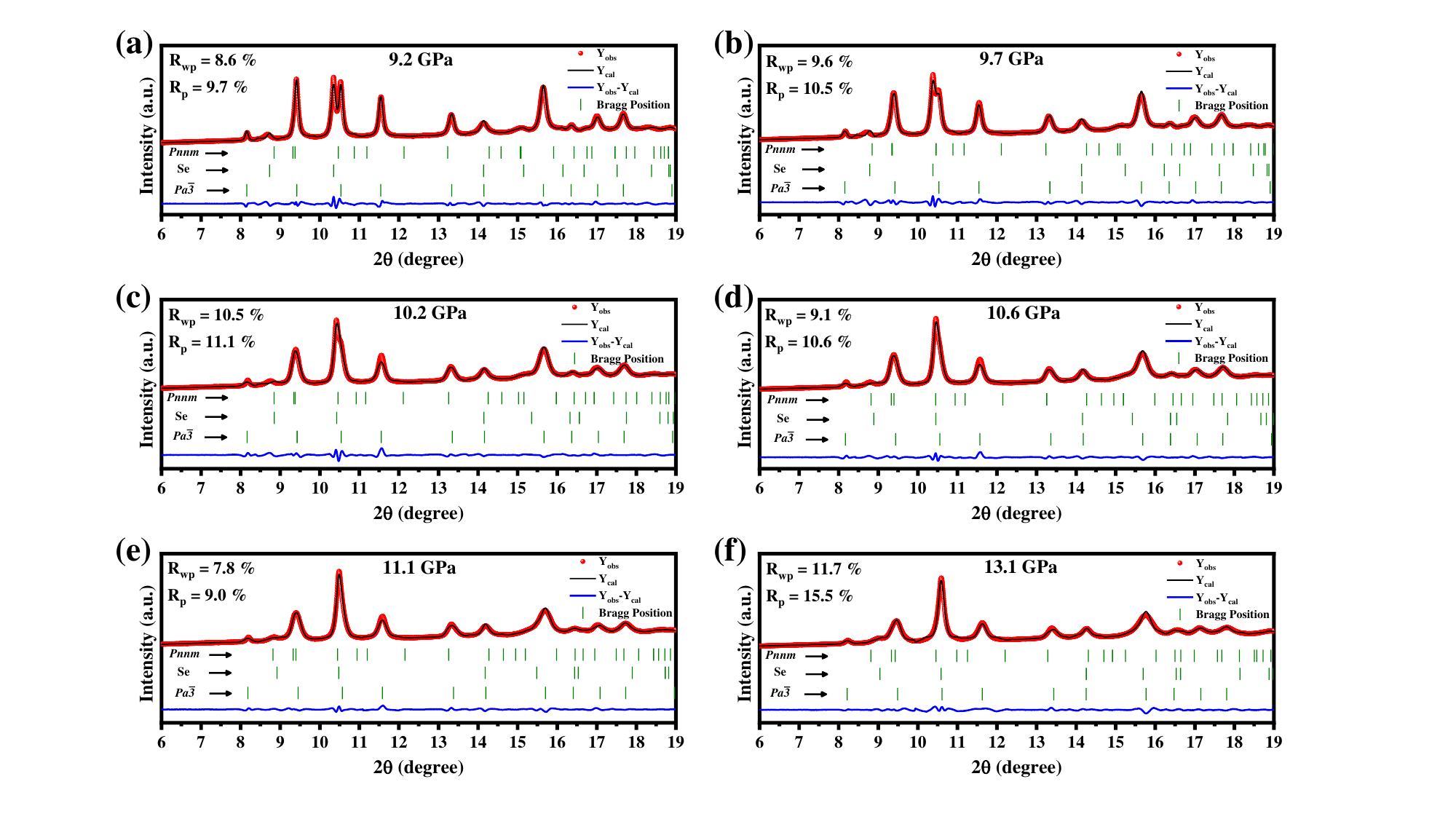}}
\renewcommand{\thefigure}{S\arabic{figure}}
\caption{(a) Rietveld refined XRD pattern at (a) 9.2 GPa, (b) 9.7 GPa, (c) 10.2 GPa, (d) 10.6 GPa, (e) 11.1 GPa, and (f) 13.1 GPa, corresponding to the phase coexistence of undistorted pyrite ($Pa\bar{3}$) and marcasite ($Pnnm$) phases of PdSe$_2$.
\label{refined4}}
\end{figure}

\begin{figure}[ht!]
\centerline{\includegraphics[scale=0.65, clip]{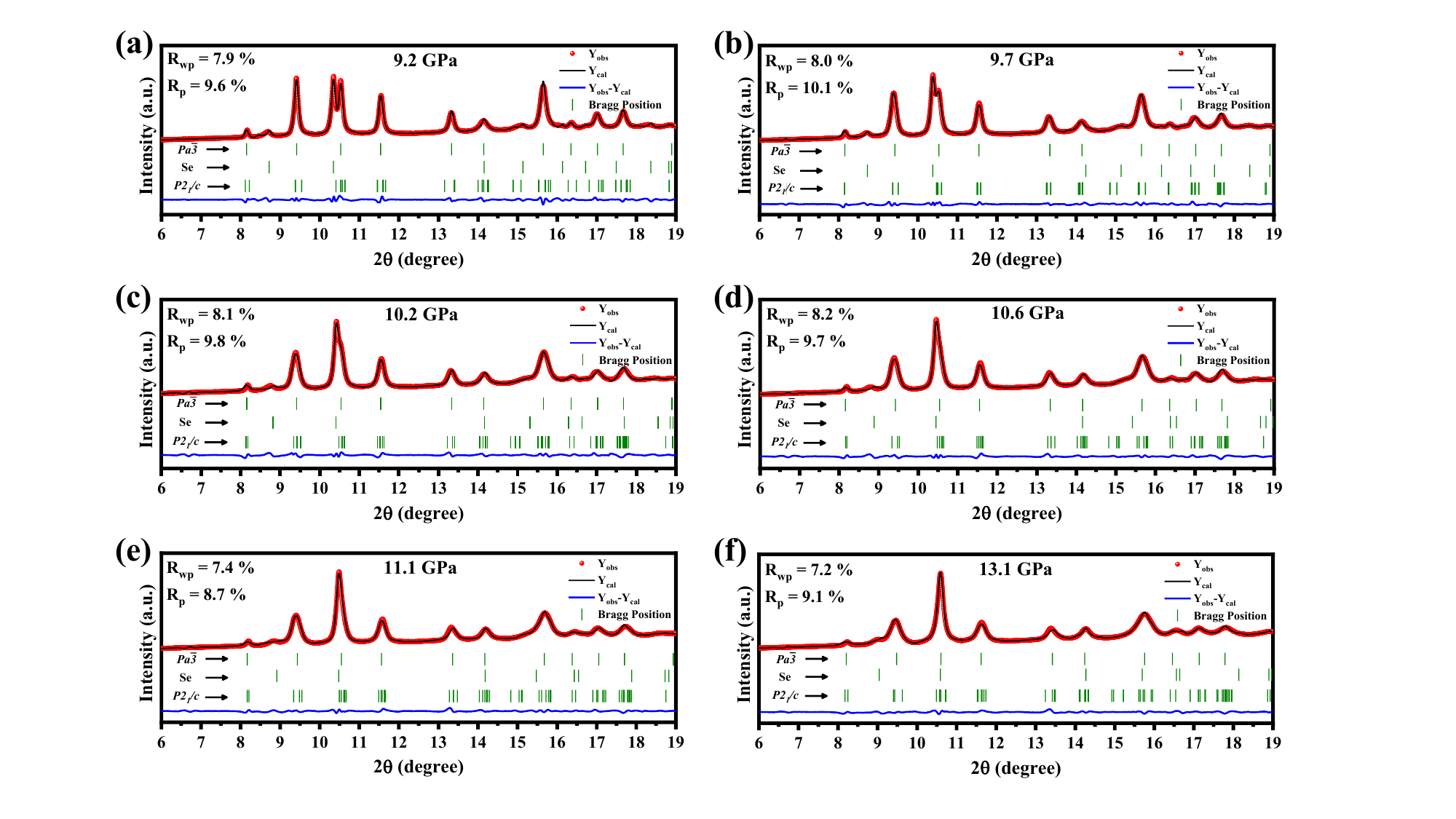}}
\renewcommand{\thefigure}{S\arabic{figure}}
\caption{(a) Rietveld refined XRD pattern at (a) 9.2 GPa, (b) 9.7 GPa, (c) 10.2 GPa, (d) 10.6 GPa, (e) 11.1 GPa, and (f) 13.1 GPa, corresponding to the phase coexistence of undistorted pyrite ($Pa\bar{3}$) and arsenopyrite ($P2_1/c$) phases of PdSe$_2$.
\label{refined5}}
\end{figure}

\begin{figure}[ht!]
\centerline{\includegraphics[scale=0.55, clip]{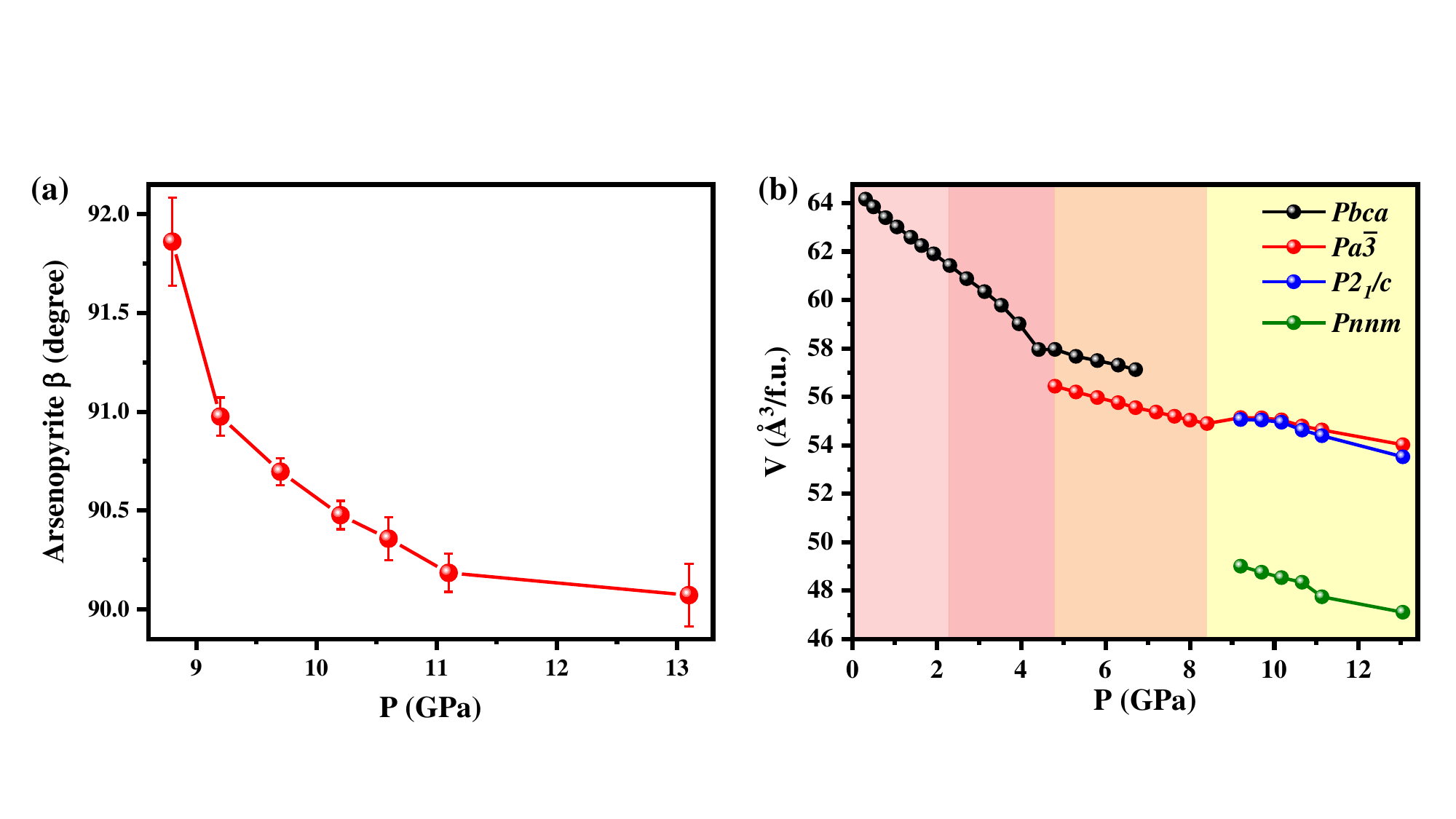}}
\renewcommand{\thefigure}{S\arabic{figure}}
\caption{(a) Variation of $\beta$ angle of the arsenopyrite ($P2_1/c$) phase under pressure. (b) Volume per formula unit as a function of pressure for the Rietveld refined four structural phases.
\label{beta}}
\end{figure}

\begin{figure}[ht!]
\centerline{\includegraphics[scale=0.55, clip]{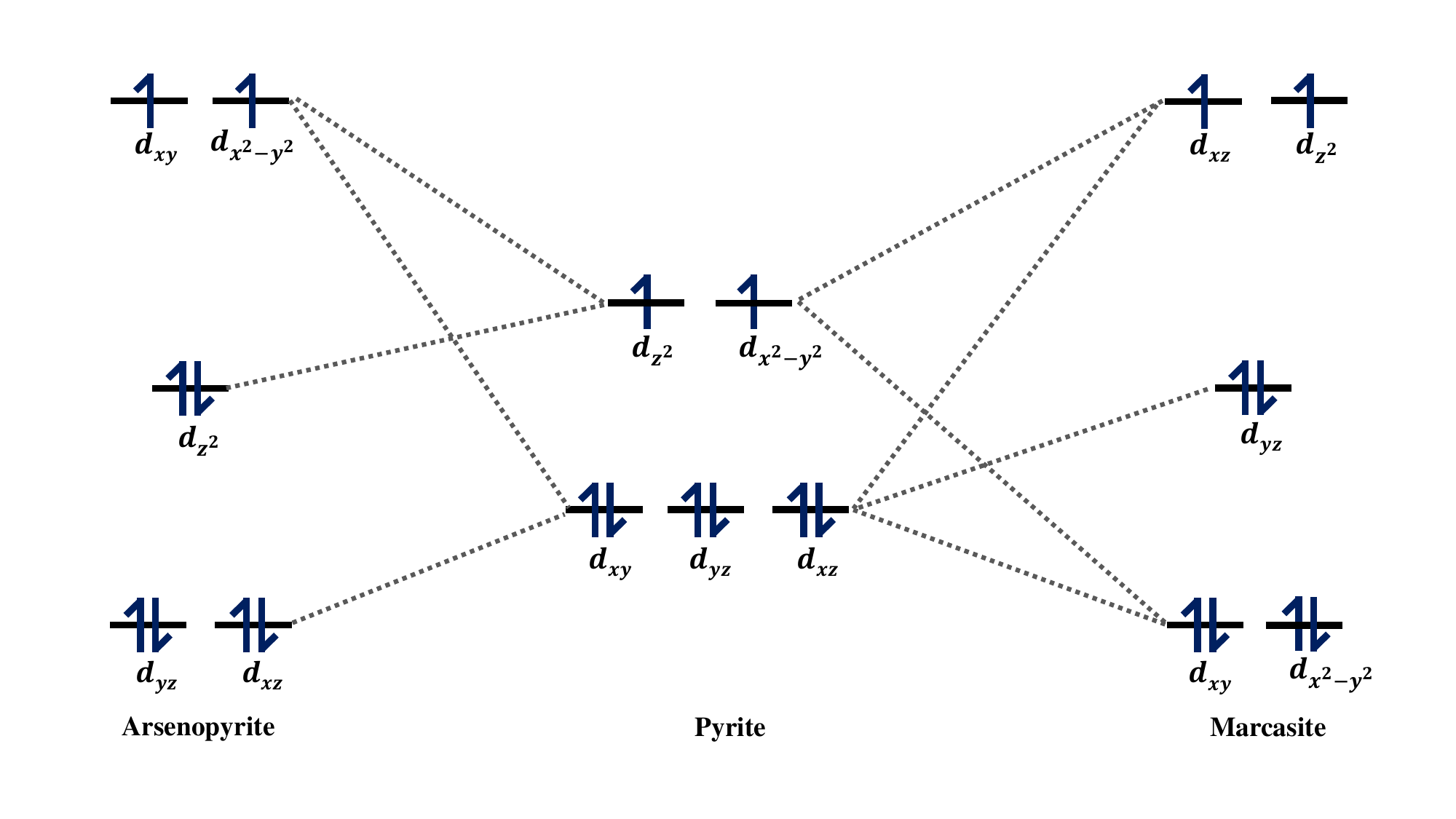}}
\renewcommand{\thefigure}{S\arabic{figure}}
\caption{Symmetric octahedral splitting for the pyrite PdSe$_2$ without any distortion. In case of both high pressure phases (marcasite and arsenopyrite), distortion appears in both $t_{2g}$ and $e_g$ orbitals and the energy ordering are schematically illustrated according to the orbital-projected band structures. 
\label{MO diagram}}
\end{figure}

\begin{figure}[ht!]
\centerline{\includegraphics[scale=0.65, clip]{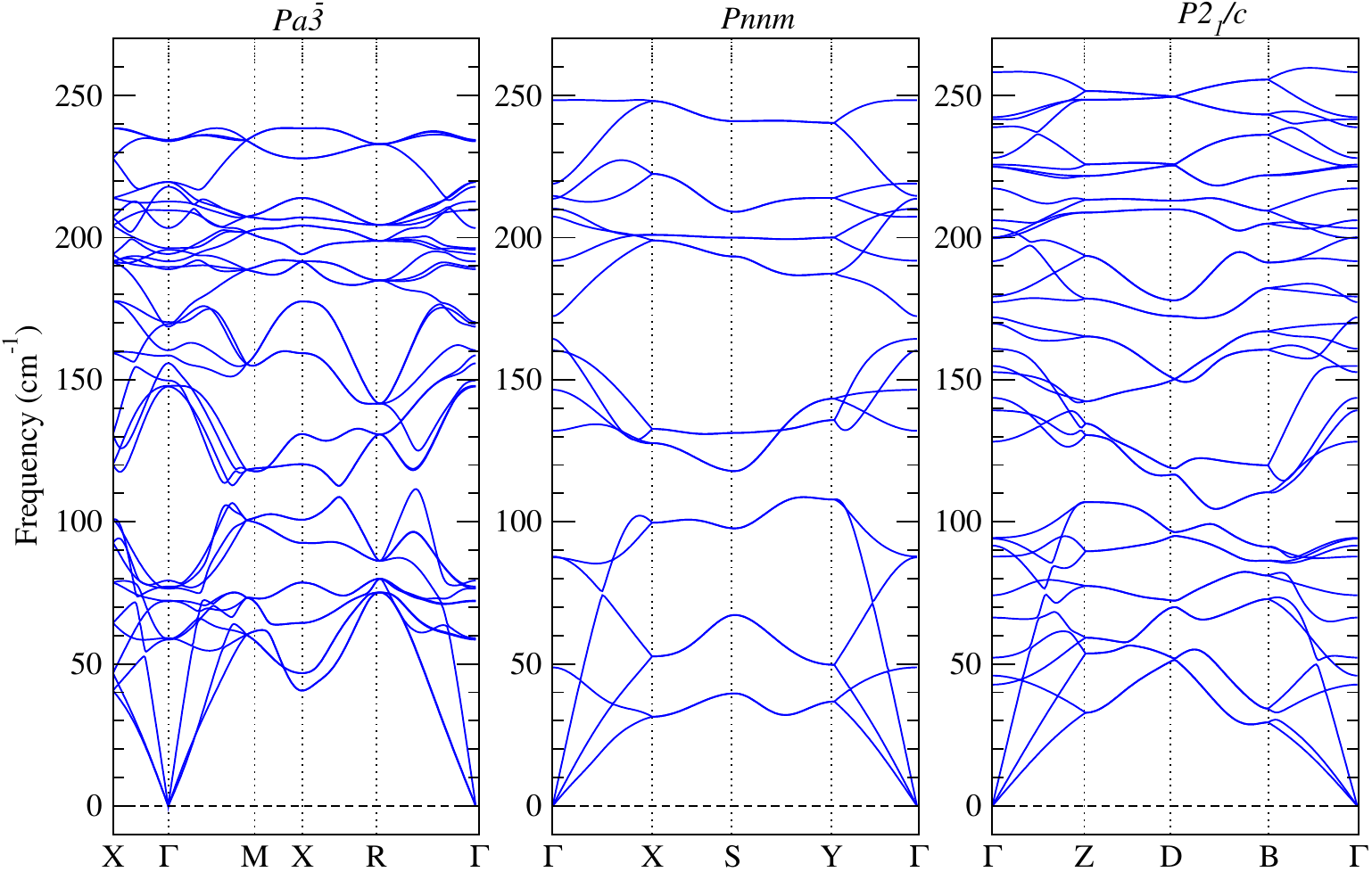}}
\renewcommand{\thefigure}{S\arabic{figure}}
\caption{Phonon band dispersions for pyrite ($Pa\bar{3}$), marcasite ($Pnnm$), and arsenopyrite ($P2_1/c$) phases at 12 GPa.
\label{phonon}}
\end{figure}

\begin{figure}[ht!]
\centerline{\includegraphics[scale=0.55, clip]{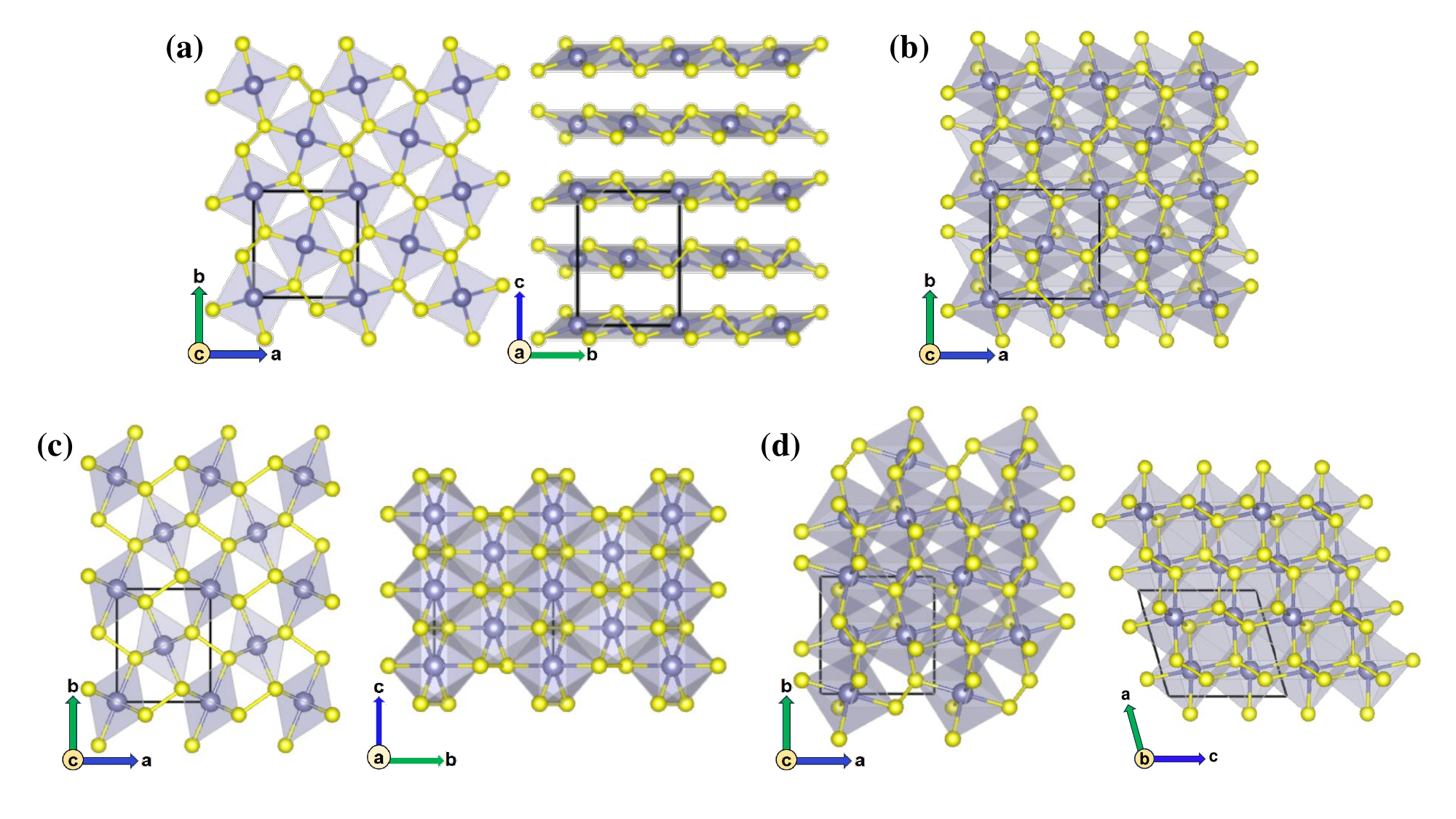}}
\renewcommand{\thefigure}{S\arabic{figure}}
\caption{Representative structures of (a) ambient ($Pbca$) phase in both top view and side view manifesting the vdW geometry with square-planar coordination, (b) non-layered pyrite phase ($Pa\bar{3}$) with symmetric octahedral geometry, (c) non-layered marcasite phase ($Pnnm$) with octahedral coordination depicted in different orientations, and (d) non-layered arsenopyrite phase ($P2_1/c$) with octahedral coordination schematized in different orientations.
\label{structures}}
\end{figure}

\begin{figure}[ht!]
\centerline{\includegraphics[scale=0.55, clip]{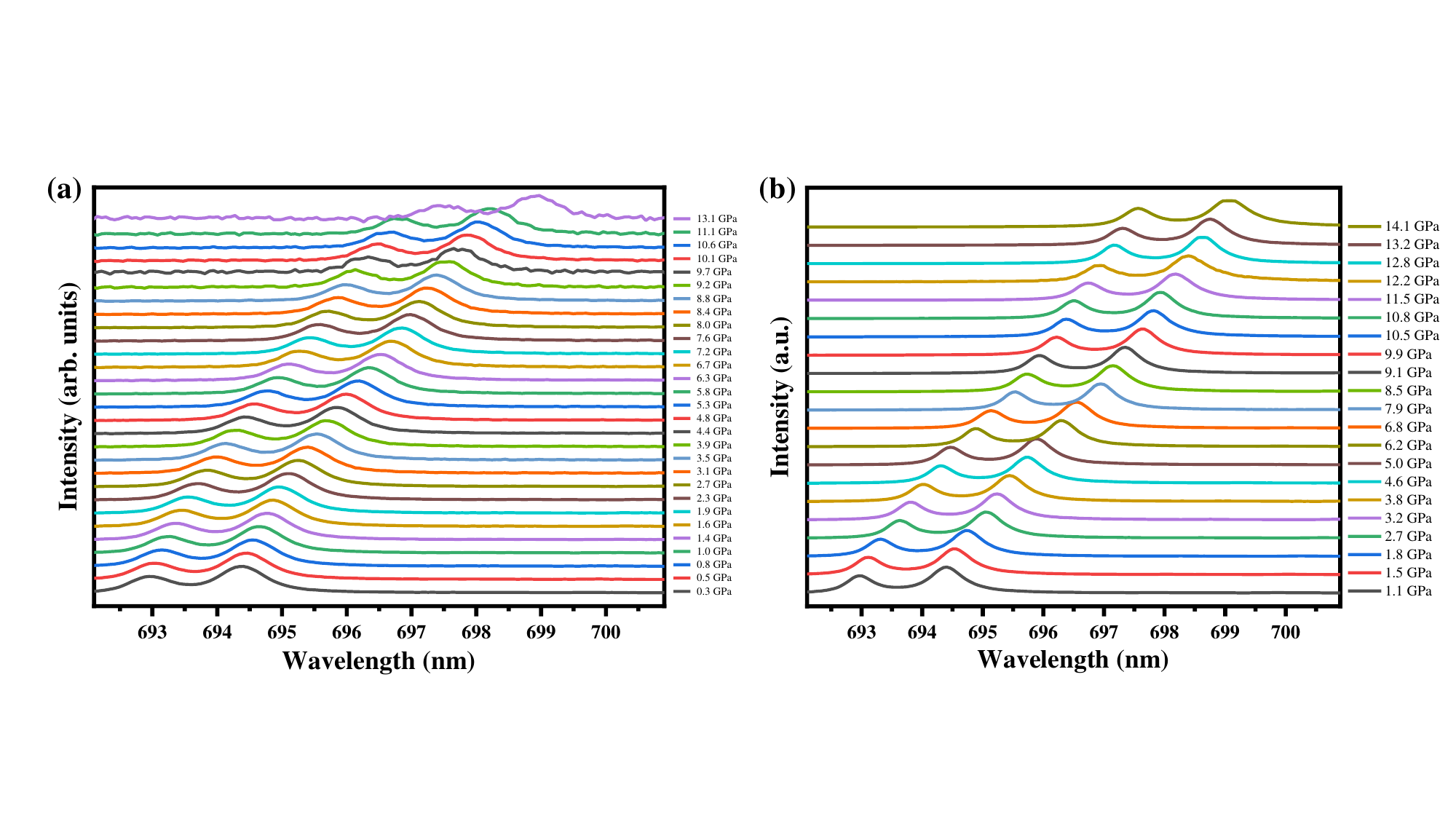}}
\renewcommand{\thefigure}{S\arabic{figure}}
\caption{Pressure-evolution of Ruby’s fluorescence spectra measured during (a) synchrotron XRD, (b) high pressure Raman spectroscopy, showing a monotonous shift with pressure.
\label{ruby}}
\end{figure}

\begin{figure}[ht!]
\centerline{\includegraphics[scale=0.55, clip]{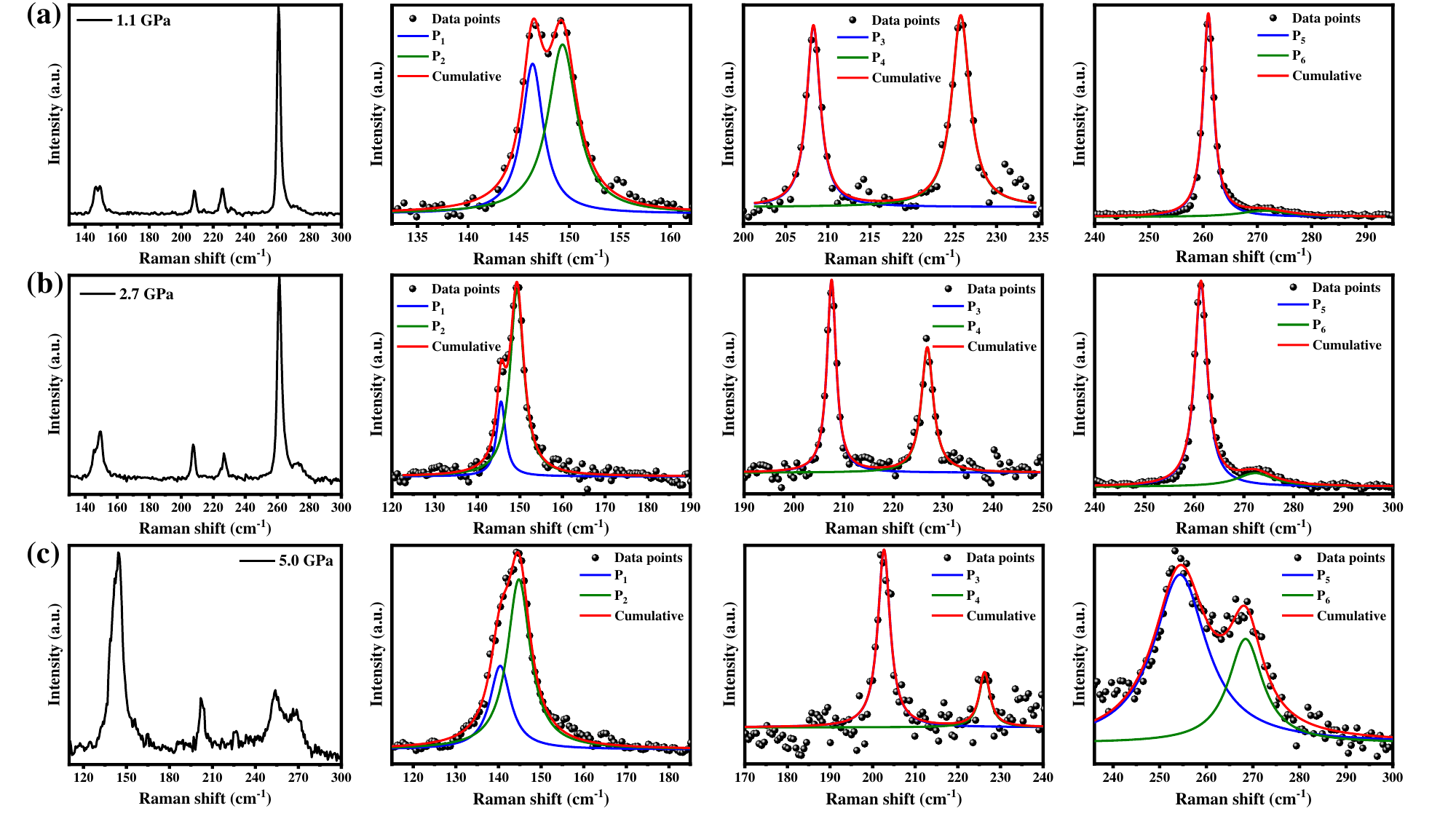}}
\renewcommand{\thefigure}{S\arabic{figure}}
\caption{Post-processing of the experimental Raman spectra at ambient condition: Experimental Raman spectrum and the fitting of characteristic Raman modes using Lorentzian function with multiple peak fitting at (a) 1.1 GPa, (b) 2.7 GPa, and (c) 5.0 GPa.
\label{Raman fitting 1}}
\end{figure}

\begin{figure}[ht!]
\centerline{\includegraphics[scale=0.55, clip]{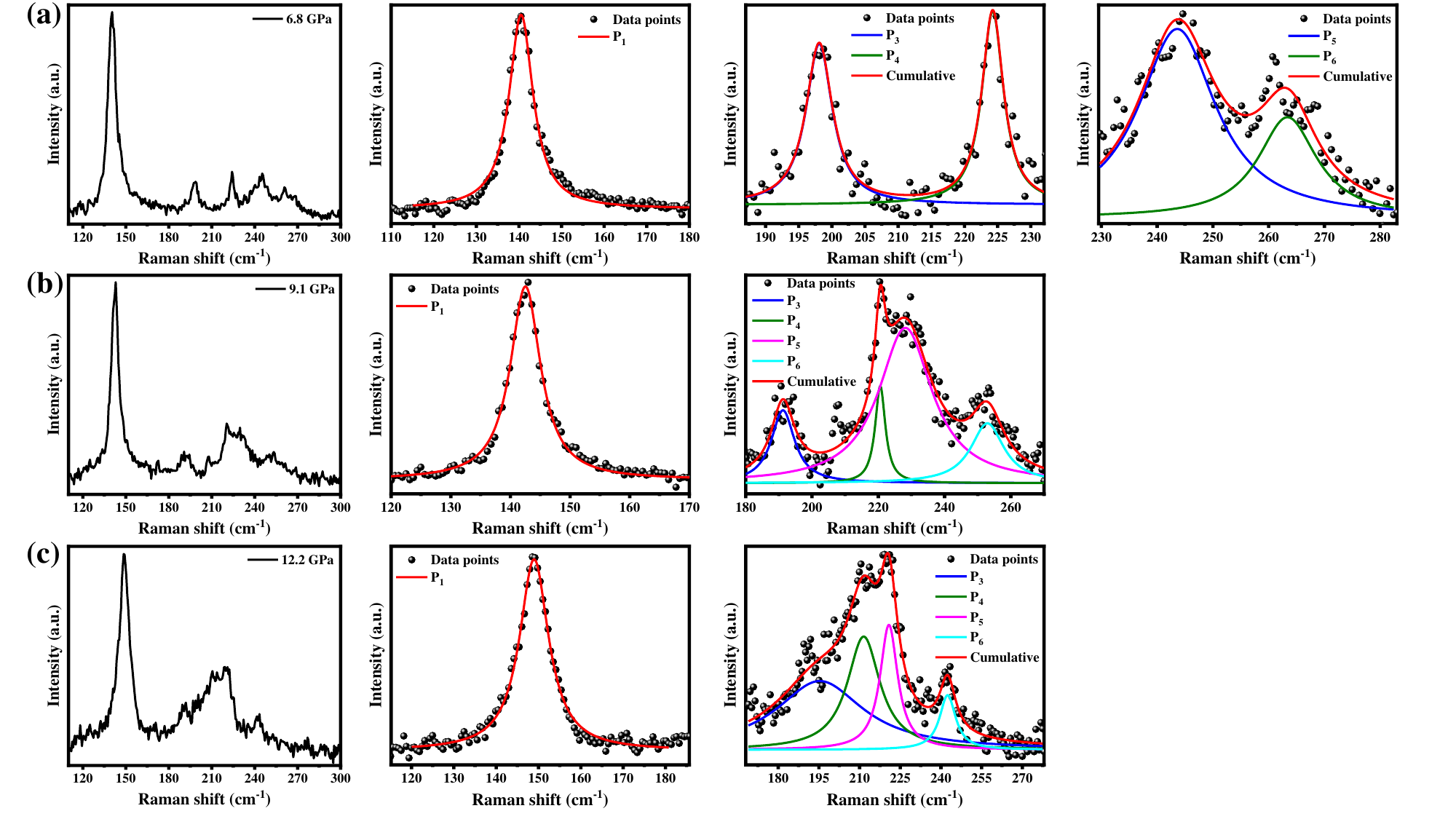}}
\renewcommand{\thefigure}{S\arabic{figure}}
\caption{Post-processing of the experimental Raman spectra at ambient condition: Experimental Raman spectrum and the fitting of characteristic Raman modes using Lorentzian function with multiple peak fitting at (a) 6.8 GPa, (b) 9.1 GPa, and (c) 12.2 GPa.
\label{Raman fitting 2}}
\end{figure}